\documentclass[aps,prb,longbibliography,twocolumn, superscriptaddress,amscd,amsmath,amssymb,verbatim]{revtex4-2}
\usepackage{graphicx,color}
\usepackage{textcomp}
\usepackage[OT1,T1]{fontenc}
\usepackage[squaren, Gray, cdot]{SIunits}
\usepackage{appendix}
\usepackage[dvipsnames]{xcolor}
\usepackage{amsmath}
\usepackage{nicefrac}
\usepackage{natbib}
\usepackage{booktabs}
\usepackage{revsymb}
\usepackage[normalem]{ulem}
\usepackage[colorlinks=true,linkcolor=blue, citecolor=blue, urlcolor=blue,
unicode=true,breaklinks]{hyperref}

\begin{document}

\title{Vortex order in magnetic frustrated GeNi$_2$O$_4$ and GeCo$_2$O$_4$ spinels}

\author{K. Beauvois} 
 \email[]{beauvois@ill.fr}
 \affiliation{CEA, IRIG, MEM, MDN, Univ. Grenoble Alpes, 38000 Grenoble, France}
 \author{J. Robert}
 \affiliation{Institut N\'eel, CNRS \& Univ. Grenoble Alpes, 38000 Grenoble, France}
 \author{M. Songvilay}
 \affiliation{Institut N\'eel, CNRS \& Univ. Grenoble Alpes, 38000 Grenoble, France}
  \author{J. Ollivier} 
\affiliation{Institut Laue-Langevin, 38000 Grenoble, France}   
 \author{B. F\aa k} 
\affiliation{Institut Laue-Langevin, 38000 Grenoble, France} 
\author{E. Ressouche} 
\affiliation{CEA, IRIG, MEM, MDN, Univ. Grenoble Alpes, 38000 Grenoble, France}
\author{N. Qureshi} 
\affiliation{Institut Laue-Langevin, 38000 Grenoble, France} 
\author{R. Ballou}
\affiliation{Institut N\'eel, CNRS \& Univ. Grenoble Alpes, 38000 Grenoble, France}  
\author{S. Petit}
\affiliation{Laboratoire L\'eon Brillouin, CEA-CNRS, Univ. Paris-Saclay, CEA-Saclay, 91191 Gif sur Yvette, France}
\author{S. Lenne}
\affiliation{Institut N\'eel, CNRS \& Univ. Grenoble Alpes, 38000 Grenoble, France} 
\author{P. Manuel}  
\affiliation{ISIS Facility, Rutherford Appleton Laboratory-STFC, Chilton, Didcot, OX11 0QX, United Kingdom}
\author{S. DeBrion}
\affiliation{Institut N\'eel, CNRS \& Univ. Grenoble Alpes, 38000 Grenoble, France}
\author{P. Strobel}
\affiliation{Institut N\'eel, CNRS \& Univ. Grenoble Alpes, 38000 Grenoble, France} 
\author{V. Simonet}
\email[]{virginie.simonet@neel.cnrs.fr}
\affiliation{Institut N\'eel, CNRS \& Univ. Grenoble Alpes, 38000 Grenoble, France}
\date{\today}

\begin{abstract}

In the search for new spin textures based on singular magnetic objects like Bloch-points or vortices, spinel compounds emerge as an interesting playground due to the interplay between magnetic anisotropy and complex interactions that extend well beyond first neighbors on a pyrochlore lattice. Based on an exploration of the exchange interaction phase diagrams of members of the Ge$B_2$O$_4$ family with $B$=Co and Ni, we show, using simultaneous modeling of inelastic neutron scattering measurements and single-crystal neutron diffraction data, that a 2-$k$ magnetic structure may be stabilized in these compounds. This leads to a short period spin vortex crystal, a variant induced by the magnetic anisotropy of the 3-$k$ Bloch-point structure predicted for isotropic spins. Our study rationalizes the formation of these multi-$k$ spin textures in frustrated antiferromagnets, as well as their anisotropy-dependent evolution.
\end{abstract}

\maketitle 

\section{Introduction}

\subsection{General interest}

 Non-collinear and non-coplanar spin textures, such as skyrmions, hedgehogs, vortex crystals, and other objects exhibiting at the continuous limit singularities of their order parameter and/or topological properties, have recently attracted considerable interest in the field of unconventional magnetism and spintronics, in particular for promising topological transport properties \cite{Gobel2021}. Among those, Bloch points, whose magnetization on a closed surface surrounding the point covers a sphere, were first identified as basic 3-dimensional (3D) defective point-like configurations in ferromagnets \cite{Feldkeller1965,Doring1968,kotiuga1989}. They are also called (anti) hedgehogs when the spins point (away) towards the singular point, thus generating an emergent magnetic field, justifying the alternative description of these Bloch points as magnetic (anti) monopoles \cite{Volovik1987,Haldane1988}. 
These objects were predicted, then observed in magnetic nano-objects like nanodots or nanowires \cite{Thiaville2003,Niedoba2005,DaCol2014}, and proposed to play a role in many topological dynamical processes occurring in spin systems such as switching of spin vortices or unwinding of skyrmions \cite{Thiaville2003,Senthil2004,Im2019,Milde2013,Schutte2014,Birch2020,Li2021,Wohlhuter2015}. 

Bloch points can also be stabilized in bulk materials in the form of 3D periodic arrangements. Crystals of hedgehogs/antihedgehogs have been proposed in itinerant cubic magnets such as the non-centrosymmetric MnGe \cite{Kanazawa2011,Kanazawa2012,Tanigaki2015,Kanazawa2016} and Mn(Si$_x$Ge$_{1-x}$) \cite{Fujishiro2019} also hosting skyrmion lattices and the centrosymmetric perovskite SrFeO$_3$ \cite{Ishitawa2020}. They originate from the combination of several helices propagating in different directions. These are examples of multi-$k$ structures (in these cases 3-$k$ and 4-$k$) i. e. built from several Fourier components of the magnetic moment distribution associated to symmetry-equivalent propagation vectors $\vec{k}$ belonging to the ordering wavevector star \cite{Park2011}. The tendency towards helimagnetism in highly symmetric crystals indeed appears to be a first step toward the realization of spin textures such as those that can occur in ferromagnets modulated by a weak Dzyaloshinskii-Moriya interaction or with additional ingredients such as long-range RKKY (Ruderman-Kittel-Kasuya-Yosida) interactions, (multiple-)spin and anisotropic interactions \cite{Yang2016,Okumura2020,Kato2023,Hayami2021}... However, multi-$k$ spin textures can also be induced in centrosymmetric insulating antiferromagnets where the key ingredient is magnetic frustration \cite{Okubo2012,Kamiya2014}.

\begin{figure}[h!]
	\includegraphics[width=1\columnwidth]{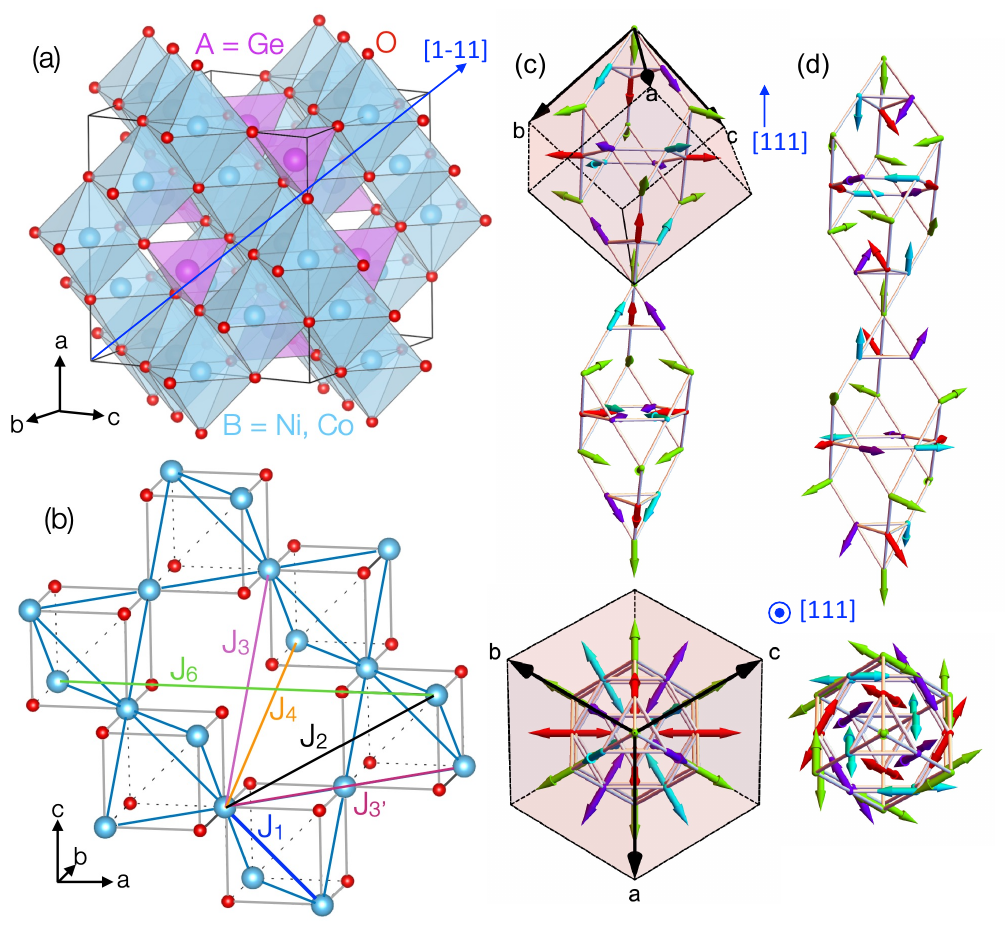}
	\caption{(a) Structure of the spinel oxide of generic formula AB$_2$O$_4$ crystallizing in the cubic space group $Fd\bar{3}m$. Arrangement of the tetrahedral site A (in purple) and of the octahedral site B (in blue). (b) Scheme of the exchange interactions up to the sixth neighbor, distinguishing the two third neighbor interactions $J_3$ and $J'_{3}$. (c-d) Sketches of the 3-$k$ cuboctahedral stack spin arrangements described in refs \cite{Lapa2012,Zhitomirsky2022}, declined in two forms from a global spin rotation: hedgehog (c) and circulating (d) Bloch-points structures. The Bloch points are elongated along a $\langle 111 \rangle$ direction (diagonal of the cube) and shown in two different views. The four Bravais lattices are indicated by different colors.}
	\label{figStructure}
\end{figure}

In this respect, spinels with the general formula $AB_2X_4$ are interesting materials. The $A$ and $B$-site cations form diamond and pyrochlore lattices, respectively, subject to magnetic frustration that can be due to lattice geometry or to competition between different interactions (see Fig. \ref{figStructure}). In this vast family of materials, complex phase diagrams \cite{Ueda2006}, $k$-degenerate ground states leading to classical or spiral spin liquids \cite{Lee2002,Gao2017} or multi-$k$ structures associated to spin textures of vortices or fractionalized skyrmion \cite{Gao2017,Gao2020} or built on the 3D coupling of triangular planes \cite{Chaix2026} were reported \cite{Tsurkan2021}. On the theoretical side, ground states of the classical Heisenberg model on a pyrochlore lattice, made of corner-sharing tetrahedra, were calculated including interactions beyond first-neighbors ($J_n$ labels the $n$th nearest neighbor interaction), which are expected to be relevant in spinel compounds (see Fig. \ref{figStructure}). In particular, exotic non-coplanar structures with a $(\frac 1 2 \frac 1 2 \frac 1 2)$ propagation vector have been predicted, featuring periodic arrays of hedgehogs \cite{Lapa2012,Aoyama2021}. 
Such a structure, also called cuboctahedral stack, is realized, for instance, through a combination of 3-$k$ components in either $J_1-J_2$ or $J_1-J_3-J_3'$ models (two distinct third neighbor interactions imposed by the spinel crystallography) \cite{Lapa2012}, or in a $J_1$-$J_6$ model \cite{Zhitomirsky2022} or even a simple antiferromagnetic $J_4$ model. This structure is shown in Fig. \ref{figStructure}c, as well as the circulating Bloch-points one in Fig. \ref{figStructure}d obtained from a global 120$^{\circ}$ rotation of the spins along the elongated axis of the structure \cite{Thiaville2003}. 
Ge spinels Ge$^{4+}B_2^{2+}$O$_4$ with $B$=Ni or Co on the pyrochlore lattice could be excellent candidates for the realization of these spin textures. However, despite decades of studies, the nature of the magnetic state and the relevant interactions involved are still unclear \cite{Diaz2006,Matsuda2008,Fabreges2017,Basu2020,Tomiyasu2011,Matsuda2011}. 
Experimentally, distinguishing a multi-$k$ structure from a single-$k$ structure remains a significant challenge since magnetic scattering from the two possible structures with equipopulated magnetic domains should be undiscernible. \cite{Schweizer2008,Herrmann1978}.

In this article, we address this question by combining neutron diffraction and inelastic neutron scattering, both analyzed using numerical calculations based on a unique Hamiltonian \cite{Gao2020,Paddison2021}. A 2-$k$ spin vortex crystal is actually deduced for both compounds, although it might still be destabilized in GeCo$_2$O$_4$ by magnetostructural effects. Interestingly, this structure can be seen as resulting from a non-trivial flattening of a 3-$k$ crystal of circulating Bloch points driven by the single-ion anisotropy whose role is highlighted as a tuning parameter.  

\subsection{State of the art on GeCo$_2$O$_4$ and GeNi$_2$O$_4$}

Although they order magnetically with the same propagation vectors $( \frac 1 2 \frac 1 2 \frac 1 2)$ \cite{Bertaut1964a,Bertaut1964b,Diaz2006}, GeNi$_2$O$_4$ (GNO) and GeCo$_2$O$_4$ (GCO) exhibit slightly different magnetic behaviors. Two successive magnetic transitions take place in GNO at $T_{\text{N}_1} = 12.1$\,K and $T_{\text{N}_2} = 11.4$\,K \cite{Crawford2003,Lashley2008} while GCO orders through a single transition at $T_\text{N} = 21$\,K but accompanied by a structural distortion from cubic to tetragonal or lower symmetry \cite{Hubsch1987,Hoshi2007,Lashley2008,Barton2014,Fabreges2017}. 
The magnetic order of both compounds was investigated by powder and single-crystal neutron diffraction. The pyrochlore structure can be described as alternating kagome and triangular layers stacked along the $\langle 111 \rangle$ directions (diagonals of the cube). Using a single-$k$ description, a collinear magnetic structure has been proposed for the two compounds consisting of antiferromagnetically stacked  kagome and triangular ferromagnetic layers \cite{Bertaut1964a,Diaz2006,Matsuda2008,Matsuda2011,Fabreges2017}. The double magnetic transition observed in GNO has been attributed to the independent ordering of the kagome and triangular layers, resulting from the zero molecular field produced by one sublattice on the other \cite{Lancaster2006,Matsuda2008}. This incidentally implies magnetic interactions beyond the first neighbors to couple all layers below $T_{\text{N}_2}$. The relevance of long-range exchange interactions up to the sixth neighbors was in fact suggested early \cite{Bertaut1964a,Plumier1967} and proved to be at play in Cr spinels \cite{Akino1971,Dwight1967,Yaresko2008}. However, in the case of the Co and Ni Ge spinels, the type and sign of the interactions and their role in the stabilization of the magnetic structure are still not established, nor is the nature of the magnetic anisotropy, since none of the proposed Hamiltonians can fully explain the proposed collinear magnetic structure. We hypothesize that these inconsistencies stem from an incorrect description of the structure as single-$k$ instead of multi-$k$. 

\begin{figure*}[t!]
	\includegraphics[width=2\columnwidth]{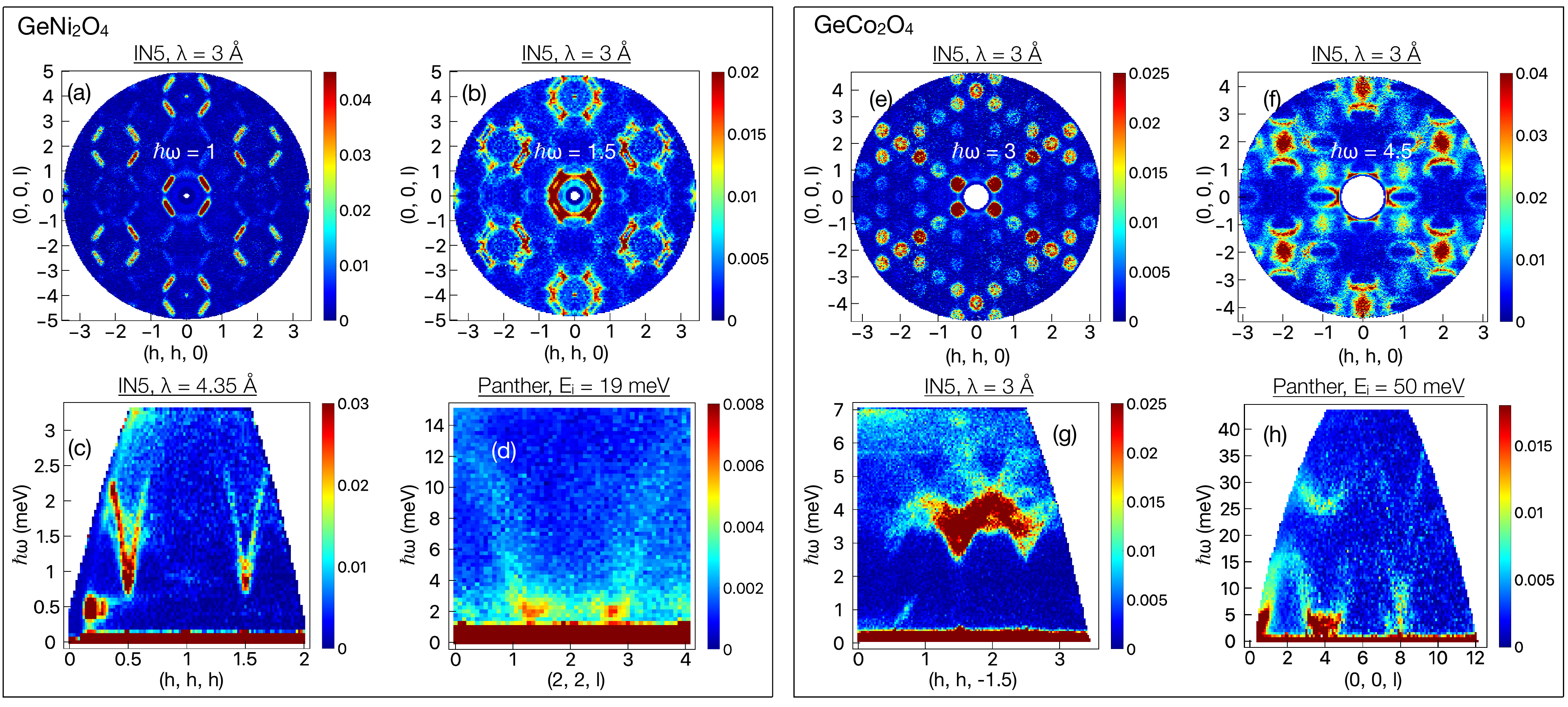}
	\caption{Left: Dynamical structure factor $S(\vec Q,\hbar \omega)$ of GNO  measured at $T=1.5$\,K. Measurements were performed on IN5 with a wavelength of 3 \AA\ (a, b) and of 4.35 \AA\ (c) and on Panther with an incident energy of 19 meV (d). (a) and (b) show constant energy cuts at $\hbar\omega=1$\,meV and $\hbar\omega=1.5$\,meV, respectively. (c) displays an $\hbar\omega$ vs ${\vec Q}$-cut along the (h, h, h) direction and (d) shows the cut along the (2, 2, l) direction, with ${\vec Q}$ expressed in r.l.u. Data were binned with a window ±0.1\,meV. Right: Dynamical structure factor $S(\vec Q,\hbar \omega)$ of GCO measured at $T=1.5$\,K  on IN5 with a wavelength of 3 \AA\ (e-g) and on Panther with an incident energy of 50 meV (h). Constant energy cuts are shown at $\hbar\omega=3$\,meV (e) and $\hbar\omega=4.5$\,meV (f). (g) displays an $\hbar\omega$ vs $\vec Q$-cut along the (h, h, -1.5) direction and (h) along (0, 0, l), with ${\vec Q}$ expressed in r.l.u. Data were binned with a window ±0.1\,meV. Note that some spurious signals are observed, e.g. in panel (c) around $\hbar\omega=0.5$\,meV at $Q = 0.2\,\text{\AA}^{-1}$ and around $\hbar\omega=3.4$\,meV at $Q = 0.6\,\text{\AA}^{-1}$.}
    \label{figOdSGNOGCO}
\end{figure*}

\section{Experimental and numerical methods}

To clarify this issue, we performed neutron scattering experiments on single-crystals. For most of our experiments, we used cylindrical single crystals grown by the floating zone method in an image furnace \cite{Hara2005}, a large one for GeCo$_2$O$_4$ (GCO) ($D = 4$\,mm and $H=3$\,cm) and a smaller one with two crystallites for GeNi$_2$O$_4$ (GNO) ($D = 4$\,mm and $H=1$\,cm). A smaller single-crystal of GCO was also grown by the flux method and used for diffraction experiments \cite{Fabreges2017}. Neutron diffraction data in GNO were obtained from single-crystal experiments on the CEA-CRG D23 two-axis diffractometer at Institut Laue-Langevin (ILL, Grenoble, France) \cite{D23}, using an incoming neutron wavelength $\lambda=1.275$\,\AA\ in an orange cryostat. Data were collected at $T=1.5$\,K. The cubic lattice parameter, determined from neutron diffraction at low temperature are $a=8.313(1)$\,\AA\, for GCO and $a=8.208(1)$\,\AA\, for GNO. The spin waves were measured in wide momentum $\vec{Q}$ and energy $E=\hbar\omega$ ranges thanks to inelastic neutron scattering (INS) experiments on the time of flight spectrometers IN5 \cite{IN5} and Panther \cite{Panther} at ILL in an orange cryostat. Both single crystals were oriented in the (hh0)-(00l) scattering plane. Several constant $\hbar \omega$ and $\vec{Q}$ maps were extracted from the data in the ordered state at the base temperature ($T=1.5$\,K) and in the correlated paramagnetic state just above $T_N$, at $T=16$\,K for GNO and at $T=30$\,K for GCO. Measurements were first performed on IN5 at several wavelengths ($\lambda=2, 3, 4.35$ \AA) and then on Panther in order to complete our measurements at higher energy, with several incident energies ($E_i=19,30,50$\,meV).  

From the numerical side, our multi-step approach was based first on spin dynamics calculations of the paramagnetic quasielastic scattering in the correlated regime, allowing, by comparison to experiments, to identify the leading terms in the Hamiltonian. The influence of each of the six exchange interactions and of the single-ion anisotropy on the periodicity of the magnetic structure was further investigated using phase diagram calculations with the Luttinger-Tisza-Bertaut method \cite{LT,Bertaut,Luttinger_1946}. Starting from the identified locations of the $(\frac 1 2 \frac 1 2 \frac 1 2)$ magnetic phase in the parameter space, mean-field minimization of the magnetic structure in the real space was performed while imposing the magnetic cell given by the propagation vector. This was coupled to calculations of the spin waves in the linear approximation to further reduce the parameter space by comparing calculations with measurements. As a final test, the Fourier components of the calculated magnetic structure were confronted to neutron diffraction Bragg peak intensities. More details are given in the Supplemental Material \cite{SupplementalMaterial}.

\section{Results}

\subsection{Magnetic excitation measurements}

For both GeNi$_2$O$_4$ (GNO) and GeCo$_2$O$_4$ (GCO), we observe complex dispersive magnetic excitations. The spectral weight is the strongest in the low energy part of the excitations above an energy gap of 0.9 and 2.8 meV for GNO and GCO, respectively (see Figs. \ref{figOdSGNOGCO}c and \ref{figOdSGNOGCO}g). This gap is in agreement with the analysis of the specific heat data and neutron spectroscopy \cite{Lashley2008}. However, we did not evidence additional gapless excitations for GNO as proposed in this reference. Starting from the intense low energy excitations rise weaker branches that strongly disperse up to $\approx 12$\,meV for GNO (Fig. \ref{figOdSGNOGCO}d) and $\approx 15$\,meV for GCO (Fig. \ref{figOdSGNOGCO}h). Additional $\hbar \omega$ vs $\vec{Q}$ scans in different reciprocal space directions are shown latter. Constant energy cuts of the low energy excitations are also displayed showing well-defined spin waves with a different shape, elongated for GNO (see Figs. \ref{figOdSGNOGCO}(a-b)) and circular for GCO (see Figs. \ref{figOdSGNOGCO}(e-f)), suggesting different ingredients in their Hamiltonian.

\begin{figure}[t!]
	\includegraphics[width=1\columnwidth]{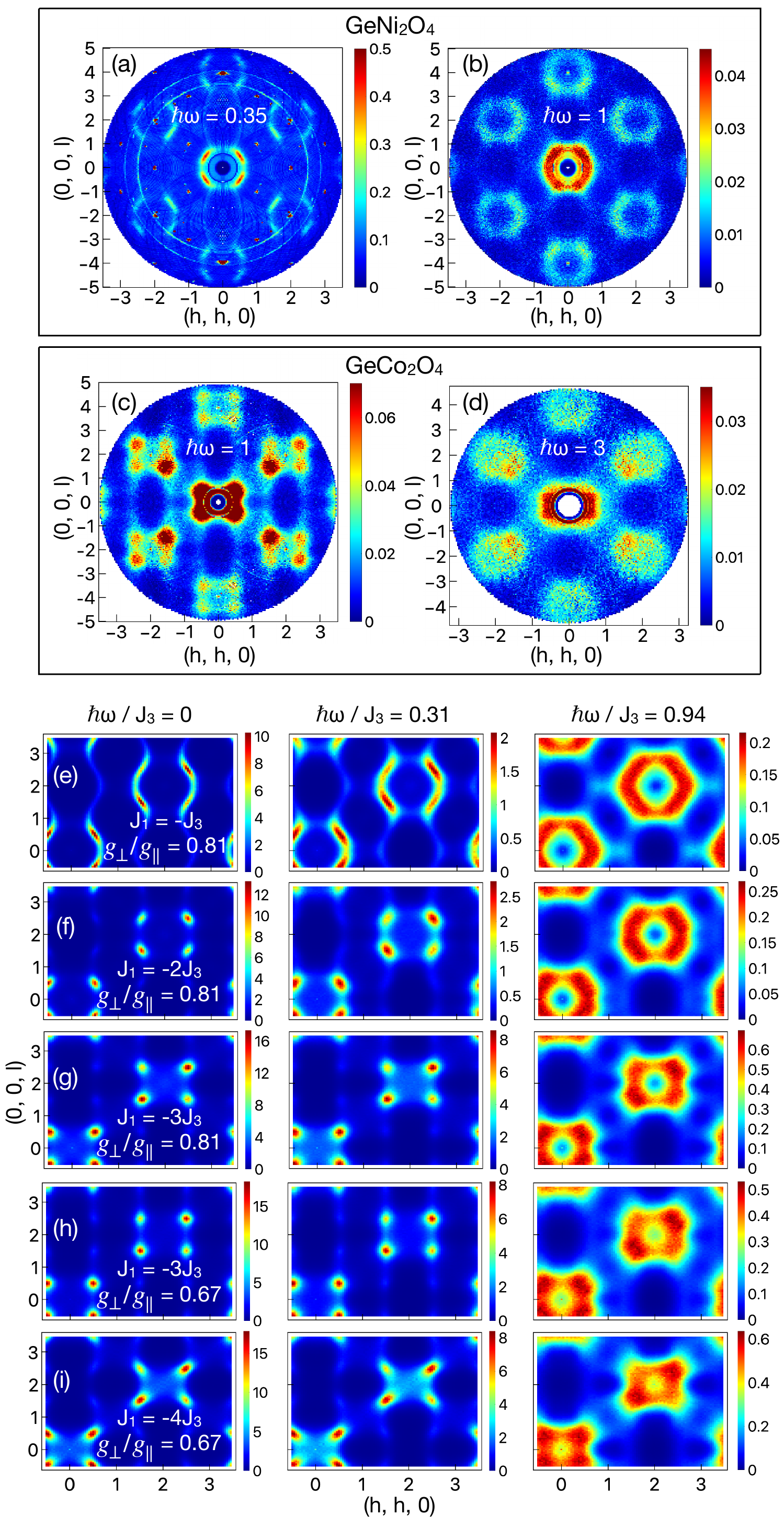}
	\caption{Measured diffuse inelastic scattering for GNO (a-b) and GCO (c-d) in the correlated paramagnetic state respectively at $T=16$\,K and $T=30$\,K with 2 energy cuts at 0.3 and 1 times the gap of the ordered state. Data collected at $T=16$\,K and $T = 30$\,K were binned using windows of ±0.1\,meV and ±0.15\,meV, respectively. Energy cuts calculated with Monte-Carlo method at {$T/T_{\text{N}}\approx 1.35$} for $J_1$/$J_3$=-1 (e), -2 (f), -3 (g) with $g_{\perp}$/$g_{\parallel}$=0.81, and for $J_1$/$J_3$=-3 (h), -4 (i) with $g_{\perp}$/$g_{\parallel}$=0.67. }
	\label{figParamagnetic}
\end{figure}

For GCO, there is an additional high energy mode at about $29$\,meV, also dispersing. The magnetic nature of this excitation is confirmed by the decrease of its intensity with increasing Q. The $^4$F electronic state of Co$^{2+}$ ions (3d$^7$) in an octahedral environment splits into two orbital triplets and one orbital singlet. With additional spin-orbit coupling, the ground orbital triplet is decomposed into one ground state spin doublet of effective spin 1/2, an excited quartet of effective spin 3/2 and an excited sextet of effective spin 5/2. This scheme can be further modified through additional trigonal or tetragonal distortion but the excitation observed at $29$\,meV corresponds to the energy gap between the effective spin $j_{\rm eff}$=1/2 doublet and effective spin $j_{\rm eff}$=3/2 levels \cite{Lines1963,Lashley2008,Sarte2019,Tomiyasu2011}. A striking feature is its dispersion, unusual for crystal field excitation. An explanation might arise from its coupling with phonons. Acoustic phonons, which cross the spin waves and the crystal field excitation, are observed at zone centers ($e.g.$ 222, 400, 444, 008, 448 ...) where the crystal field modulation is the strongest (see Fig. \ref{figOdSGNOGCO}h). An alternative explanation is that this modulation results from significant mixing between the $j_{\rm eff}$ spin-orbit levels due to strong exchange constants as proposed in other Co-based compounds \cite{Sarte2019,Elliot2021}.

In GNO, the lowest $^3$F state of Ni$^{2+}$ ions (3d$^8$) ions also splits in the cubic crystal field but the orbital singlet is the lowest energy one, leading to a zero orbital moment and spin triplet of value 1. Anisotropy can arise from a second-order effect implying a coupling of the ground state with the upper states at much higher energies when combining trigonal distortion and spin-orbit coupling. The ground spin triplet is further split and the doublet is usually lower in energy \cite{Lines1963,Crawford2003}. In this case, no crystal field transition is expected in the energy range investigated. 

With the exception of the $j_{\rm eff}$=3/2 high energy excitation in GCO, we attribute the magnetic excitations observed in both compounds to spin waves, as they exhibit clear dispersion and because our data can be modeled in the linear spin wave approximation using an adequate Hamiltonian, including single-ion anisotropies, as detailed in the next sections. Our results therefore shed new light on the nature of these excitations compared to the previously proposed interpretation in terms of confined excitations resulting from clustering \cite{Tomiyasu2011}. 

\subsection{Diffuse scattering and phase diagrams}

An inspection of the crystallographic structure gives an indication about the relevant Heisenberg exchange interactions $J_{ij}$ that enter the effective Hamiltonian of the spinels (see details in the Supplemental Material \cite{SupplementalMaterial}):
\begin{equation}
\mathcal{H}=-\sum_{ij} J_{ij} \ \mathbf{S}^T_i \ \tilde g^T_i \ \tilde g_j \ \mathbf{S}_j,
\label{eqham}
\end{equation}
where $\mathbf{S}_i$ is a spin operator at site $i$, $J_{ij}=J_1-J_6$ are the isotropic interactions between pairs of spins $i$ and $j$ up to the sixth neighbors (see Fig. \ref{figStructure}b), and $\tilde g_i$ is the g-tensor at site $i$. The latter comprises two components, parallel $g_{\parallel}$ and perpendicular $g_{\perp}$ to the local $\langle 111 \rangle$ three-fold axis at the Co/Ni site \cite{Yan2017} reflecting the single-ion anisotropy. 

We considered interactions corresponding to Ni-Ni/Co-Co bond distances of 2.90/2.94 ($J_1$), 5.03/5.09 ($J_2$), 5.81/5.88 ($J_3$ and $J'_3$), 6.49/6.57 ($J_4$) and 8.21/8.31 ($J_6$) \AA. There are two contributions to the first neighbor interaction, a positive ferromagnetic (FM) direct exchange which concerns only GCO and a superexchange interaction through the O anions which is also FM according to the Goodenough-Kanamori rules \cite{Goodenough} for a superexchange angle equal to 90$^{\circ}$. $J_1$ is therefore expected to be FM and stronger for GCO. Although associated to the longest distance, $J_6$ could still be relevant since the super-superexchange path is a direct 180$^{\circ}$ one leading to a negative antiferromagnetic (AFM) interaction. The sign and strength of the other interactions are difficult to anticipate since they can depend, not only on the distance and path angles, but also on the degree of covalency of the Ge ions that can contribute to some of the exchange paths. Note for instance that there are two distinct third neighbor interactions $J'_{3}$ and $J_3$ with different super-superexchange paths, including an interstitial magnetic ion and 90$^{\circ}$ paths for the former and 125$^{\circ}$ paths mediated by a Ge for the latter. They are therefore expected to be very different and stronger for $J_3$ \cite{Plumier1966}. It has been proposed in the literature that $J_2$ should be weak while $J_3$ and $J_4$ could be crucial in stabilizing the structure \cite{Matsuda2008}.

\begin{figure}[t!]
	\includegraphics[width=1\columnwidth]{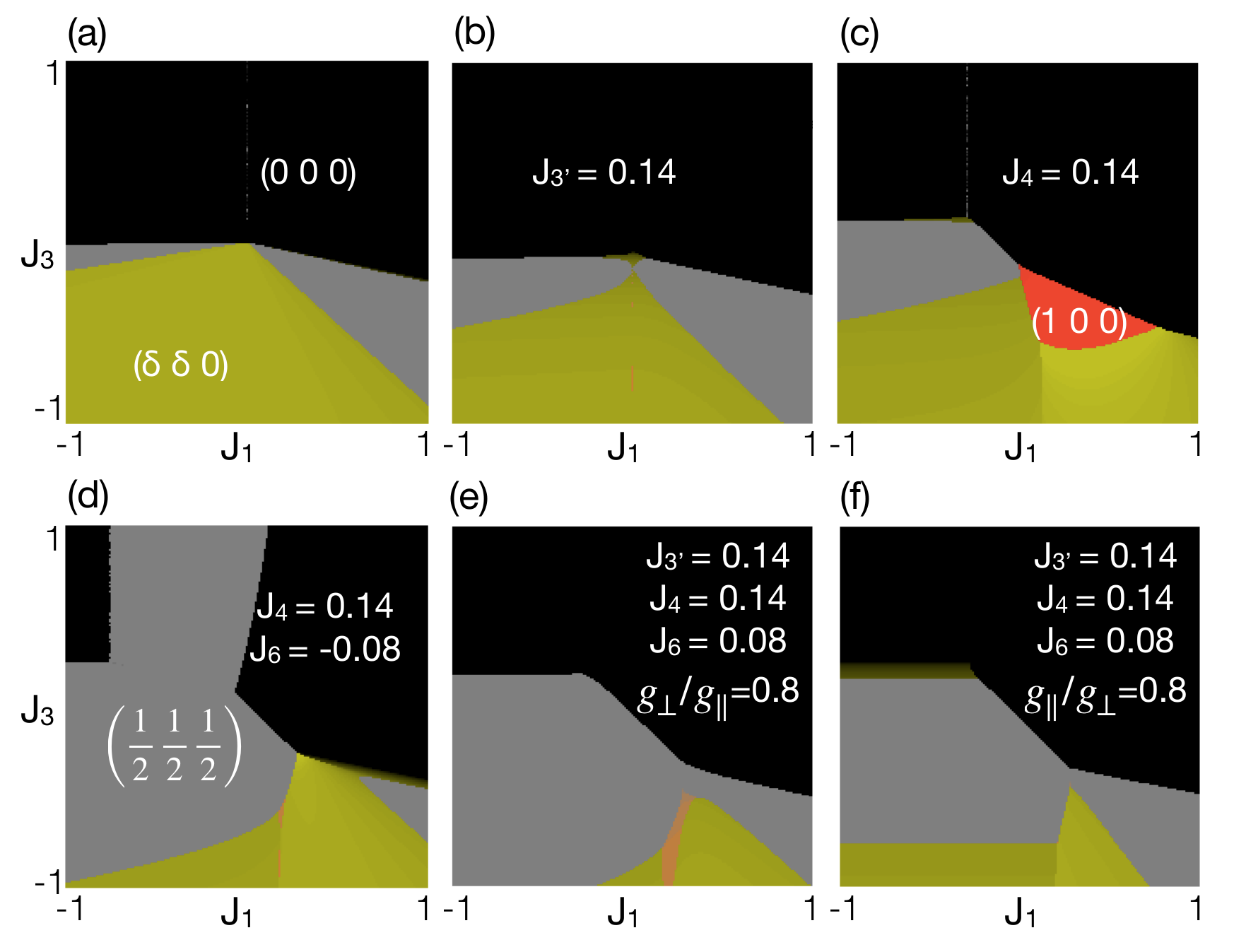}
	\caption{$J_1$/$J_3$ phase diagrams for the pyrochlore lattice showing the stability region of the different ordered phases characterized by their propagation vector (000), ($\delta\delta 0$), (100) and $(\frac 1 2 \frac 1 2 \frac 1 2)$, the one observed in GNO and GCO (in grey) (a). Modification of this phase diagram when including $J'_{3}$ (b), $J_4$ (c), $J_6$ (d), $J'_{3}$-$J_4$-$J_6$ with a single-ion Ising anisotropy $g_{\perp} < g_{\parallel}$ (e) and an easy-plane one $g_{\parallel} < g_{\perp}$ (f).}
	\label{figPhaseDiagram}
\end{figure}

To get further hint on the hierarchy of the exchange couplings, we explore the inelastic diffuse scattering in the paramagnetic state at $\approx$1.35 $T_\text{N}$. Constant-energy cuts are shown for GNO in Fig. \ref{figParamagnetic}(a-b) and GCO in Fig. \ref{figParamagnetic}(c-d) at $\approx$1/3 and $\approx$1 times the value of the ordered state gap, which collapses above $T_\text{N}$. The signal is still well-structured at low energy revealing that significant correlations persist in the paramagnetic state in agreement with previous results \cite{Lashley2008,Tomiyasu2011}. The low energy signal consists in four spots rising from the magnetic satellites around some zone centers. They are elongated for GNO and  for GCO. At higher energies, the signal becomes more diffuse. Monte-Carlo calculations using a model for Ising-like spins with ferromagnetic $J_1$ and antiferromagnetic $J_3$ are shown in Fig. \ref{figParamagnetic}(e-i) at different energies and increasing $|J_1$/$J_3|$ ratio. At $|J_1$/$J_3|$=1, the calculated four spots are elongated as in GNO. When the ratio $|J_1$/$J_3|$ increases, they become isotropic as in GCO ($|J_1$/$J_3|\approx$3) before becoming elongated in a 90$^{\circ}$ direction for higher ratio. Note that the elongation of the spots and their direction actually reflects the evolution of the spin pair correlations at the proximity of neighboring phases, ($\delta\delta 0$) for unitary ratio and (000) for larger ratio (see Fig. \ref{figPhaseDiagram}a). The shape of the spots is also influenced by the single-ion anisotropy as seen in Fig. \ref{figParamagnetic}(g-h). This simple FM $J_1$ - AFM $J_3$ model is therefore able to reproduce rather well the paramagnetic features of both compounds, whose difference is due to a larger $|J_1$/$J_3|$ ratio for GCO than for GNO. These two interactions must be the strongest and their influence subsists above the ordering temperatures while weaker ones are averaged by thermal fluctuations in this temperature range. 

\begin{figure*}[t!]
\includegraphics[width=2\columnwidth]{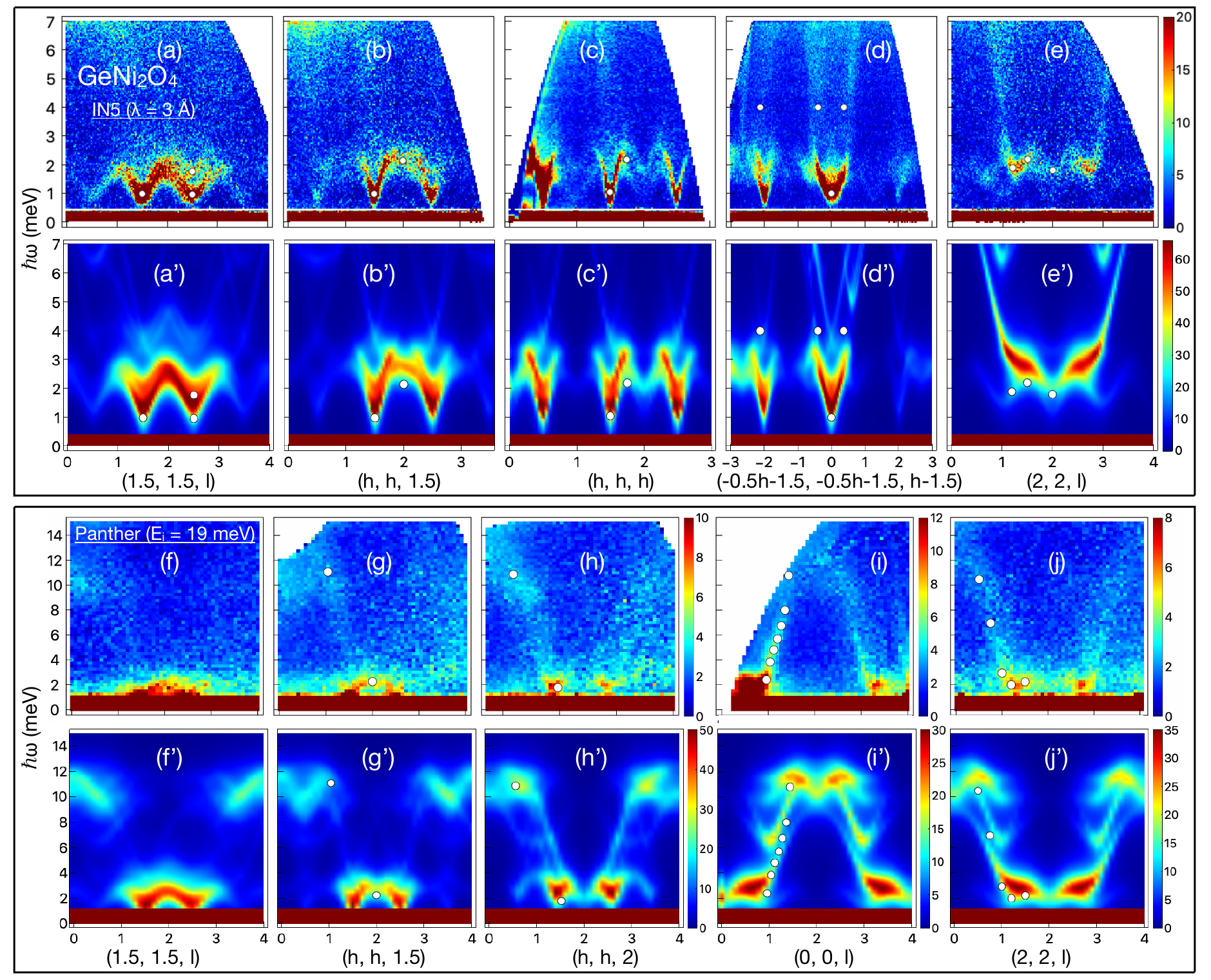}
\caption{Measured (above on IN5 and below on Panther) and calculated dynamical structure factor $S(\vec Q,\hbar \omega)$ in GNO using the exchange constants and the single-ion anisotropy of Table \ref{Table1}: $\hbar\omega$ vs $\vec Q$-cuts (r.l.u) along several directions of the reciprocal space. The width of the calculated excitations was taken as the energy resolution. The white points on top of the experiments and calculations give the fitted energy positions of the experimental spin wave dispersion (see Supplemental Material \cite{SupplementalMaterial}). }
\label{figCalcMeasOdSGNO}
\end{figure*}

\begin{figure*}[t!]
	\includegraphics[width=2\columnwidth]{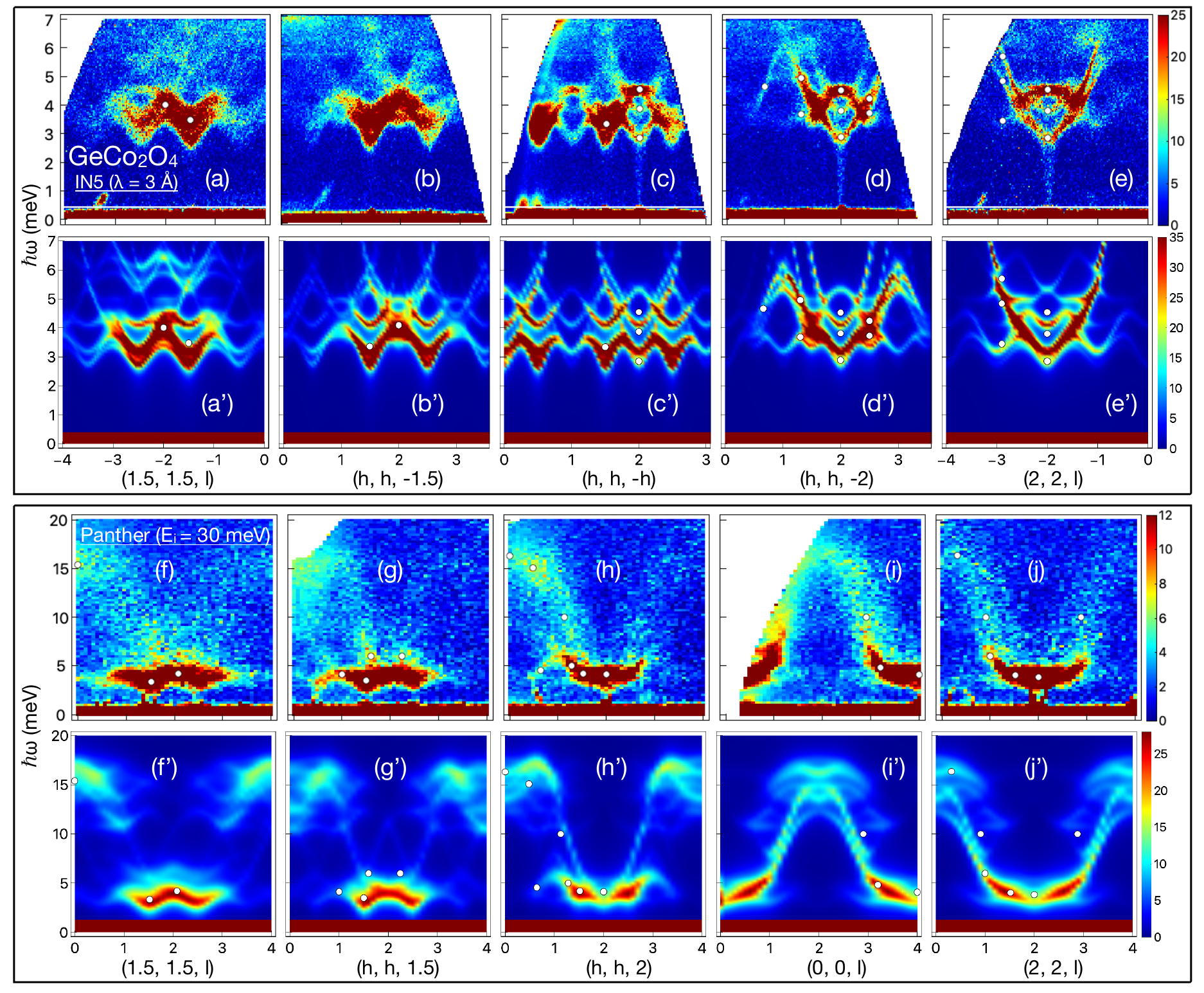}
	\caption{Measured (above on IN5 and below on Panther) and calculated dynamic structure factor $S(\vec Q,\hbar \omega)$ in GCO using the exchange constants and the single-ion anisotropy of Table \ref{Table1}: $\hbar\omega$ vs $\vec Q$-cuts (r.l.u) along several directions of the reciprocal space. The width of the calculated excitations was taken as the energy resolution. The white points on top of the experiments and calculations give the fitted energy positions of the experimental spin wave dispersion (see Supplemental Material \cite{SupplementalMaterial}). }
	\label{figCalcMeasOdSGCO}
\end{figure*}

Starting from this $J_1$-$J_3$ model, we then investigated the classical phase diagram using the Luttinger-Tisza-Bertaut approximation (see Appendix E of Supplemental Material \cite{SupplementalMaterial} for more details about the method) and including the additional interactions. We first checked that a FM $J_1$ and an AFM $J_3$ stabilize a $(\frac 1 2 \frac 1 2 \frac 1 2)$ propagation vector, which is the one observed in both compounds. This indeed corresponds to the grey region in the lower right corner in panel (a) of Fig. \ref{figPhaseDiagram}. Note that we checked that the parameters for the other grey $(\frac 1 2 \frac 1 2 \frac 1 2)$ regions do not reproduce our measured spin waves. To be in the correct phase, our calculations indicate that $J_3$ and $J'_3$ must have an opposite sign. Therefore, $J'_{3}$ must be FM, in which case it extends the $(\frac 1 2 \frac 1 2 \frac 1 2)$ region (see panel b). $J_4$ is expected to be AFM or weakly FM to favor the $(\frac 1 2 \frac 1 2 \frac 1 2)$ region. An AFM $J_6$ interaction has a very strong effect in increasing the stability region of the $(\frac 1 2 \frac 1 2 \frac 1 2)$ phase while a FM one kills it. For an unfavorable FM $J_4$, $J_6$ allows to retrieve the phase of interest (panel d). The single-ion anisotropy extends or reduces slightly the grey phase of interest when changing from easy-plane to Ising (panels e-f). This analysis gives some hints about the ingredients necessary to favor a $(\frac 1 2 \frac 1 2 \frac 1 2)$ propagation vector. 

\subsection{Spin wave modeling}

We then compared our measured magnetic excitations with calculations based on the linear spin wave theory, using the Hamiltonian of equation (\ref{eqham}). To do this, we first allowed the magnetic structure to converge to the state that minimizes the energy in equation (1), while imposing the periodicity given by the Luttinger-Tisza-Bertaut method. The magnetic structure thus obtained (discussed in more detail later) then served as a starting point for constructing the excitation spectrum. For both GNO and GCO, several sets of exchange parameters allow to reproduce rather well the measured spin waves. Note that for GCO, this modeling excludes the high energy mode at $29$\,meV. We then added two constraints: {\it (i)} the hierarchy of the $J_1$ and $J_3$ interactions obtained from the comparison of the measured versus calculated paramagnetic diffuse scattering (see Fig. \ref{figParamagnetic}) and {\it (ii)} a stabilized ordered phase with the observed $(\frac 1 2 \frac 1 2 \frac 1 2)$ propagation vector from the Luttinger-Tisza-Bertaut computation of the phase diagrams. The best sets of parameters using this methodology are reported in Table \ref{Table1} and the resulting spin waves are shown in Figs. \ref{figCalcMeasOdSGNO} for GNO and \ref{figCalcMeasOdSGCO} for GCO. All the magnetic domains were included in the calculations. The strongest interactions are a FM $J_1$ and an AFM $J_3$ as already hinted by the paramagnetic diffuse scattering analysis, with a larger $J_1$ and a larger $|J_1$/$J_3|$ ratio in GCO (2.9) compared to GNO (1.15). The Ising-like anisotropy is also stronger for GCO than for GNO. 

Noteworthy, this minimal Hamiltonian is already relevant to account for the overall shape and the weak branches rising up to a maximum of 12 (15) meV for GNO (GCO) but not to reproduce the details of the intense signal at low energy. Additional FM $J_4$ and $J'_{3}$ and AFM $J_2$ and $J_{6}$ interactions provide the necessary ingredients for this behavior (see Fig. SM3 in the Appendix D of the Supplemental Material \cite{SupplementalMaterial}). The signs of these weakest exchange interactions $J_2$, $J_3'$, $J_4$, and $J_6$ are uniquely determined, while their amplitudes are determined relative to $J_1$ and $J_3$, which are constrained by the paramagnetic diffuse scattering. Interestingly, models with only $J_1$, $J_3$ and $J_4$ can already reproduce the spin wave spectra reasonably well, highlighting the essential role of a ferromagnetic $J_4$ in accurately describing the low energy spin wave modes. The small $J_6$ term seems also necessary to keep the system in the $(\frac 1 2 \frac 1 2 \frac 1 2 )$ phase while maintaining a good comparison between the calculated and measured spin waves. Note that for GCO, some additional spectral weight is observed in the experimental data compared to the calculations at zone centers where phonons cross the spin waves. We suspect that this discrepancy can be due to some coupling of both types of excitations, in line with the sensitivity of the compound to magnetoelastic effects \cite{Watanabe2011,Fabreges2017,Pramanik2021}. Moreover, if there is a mixing between the wavefunctions of the ground state doublet and of higher-energy multiplets due to overly strong interactions (responsible for the dispersion of the spin–orbit exciton), linear spin wave theory is no longer strictly valid; nevertheless, it can still reproduce the low-energy data as a first approximation. The differences could also be due to anisotropic interactions that have been proposed theoretically to be significant in cobaltates  \cite{Liu2018,Sano2018}. For comparison, we also computed the spin waves using the model proposed by Pramanik {\it et al.} from density functional theory calculations confronted to Infrared and Raman spectroscopy \cite{Pramanik2021} (see Table \ref{Table1}). The agreement is clearly much less satisfactory (see Fig. SM4 in the Supplemental Material \cite{SupplementalMaterial}).  

\begin{table}[!h]
	\centering
	\begin{tabular}{|p{1.4cm}||p{0.7cm}|p{0.7cm}|p{0.7cm}|p{0.9cm}|p{0.7cm}|p{0.7cm}|p{0.9cm}|p{0.9cm}|}
		\hline	 & \textcolor{blue}{$J_1$}  & $J_2$ & \textcolor{magenta}{$J_3$} & $J_1 / J_3$ & \textcolor{purple}{$J_{3'}$} & \textcolor{orange}{$J_4$} & \textcolor{green}{$J_6$} & g$_{\perp}$/g$_{\parallel}$  \\
		\hline 	 GNO                     & 1.95 & 0    & -1.7 & -1.15 & 0.27 & 0.27  & -0.101 & 0.81 \\
		\hline   GCO	                 & 4  & -0.65    & -1.4 & -2.9  & 0.35  & 0.4   & -0.192  & 0.65 \\
		\hline   GCO \cite{Pramanik2021}	 & 3.9  & -0.7 & -2.0 & -1.95 & -0.4 & 0     & 0      & 0.65 \\
		\hline
	\end{tabular}
	\caption{Values of the exchange interactions (meV) and of the magnetocrystalline anisotropy of GNO and GCO deduced from the best comparison between our spin waves calculations and the INS measurements while stabilizing a $(\frac 1 2 \frac 1 2 \frac 1 2)$ propagation vector (see Figs. \ref{figCalcMeasOdSGNO} and \ref{figCalcMeasOdSGCO}). Positive (negative) values represent FM (AFM) magnetic interactions. In the last line of the Table are also given the interaction parameters calculated by Pramanik {\it et al.} \cite{Pramanik2021} for GCO, which yield a propagation vector $(0.69, 0.69, 0)$.}
	\label{Table1}
\end{table}

\begin{figure}[h!]
	\includegraphics[width=1.0\columnwidth]{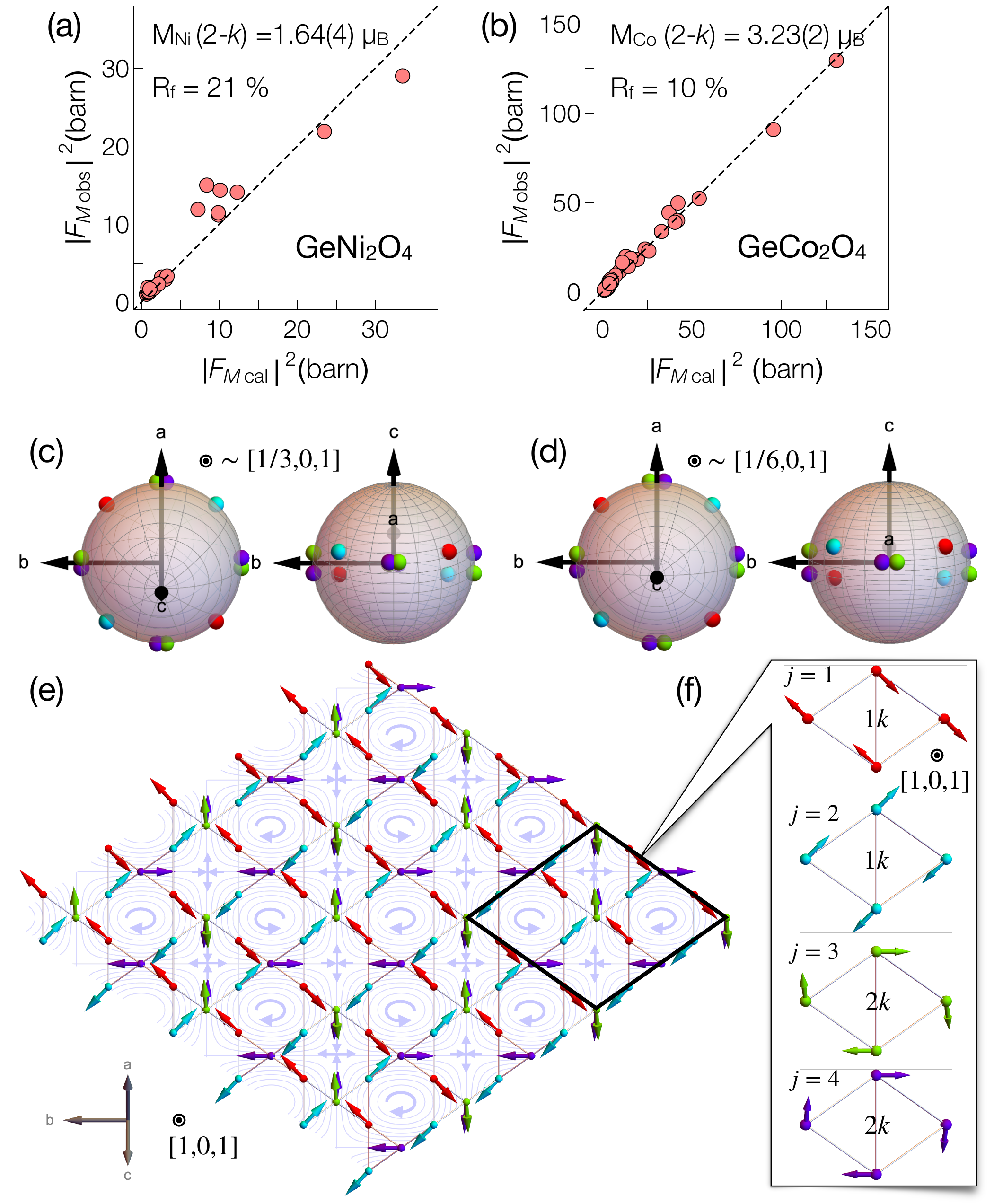}
	\caption{Measured magnetic Bragg peak intensities versus calculated ones from the models of Table 1 for GNO (a) and GCO (b). The refined magnetic moment is given as well as the goodness of the fit R$_{\rm f}$. Distribution of the spin orientations for GNO (c) and GCO (d) for the magnetic domain based on the Fourier components of  $(\frac 1 2 \frac 1 2 \frac 1 2)$ and $(\frac 1 2\frac {\bar 1} {2}\frac 1 2)$ wavevectors: The extremities of the spins are shown as balls on a sphere, viewed perpendicular and along the $(hkl)$ direction indicated in black to emphasize the almost-2D character of the magnetic order. The colors correspond to the four Bravais lattices. (e) Corresponding spin arrangement for GNO (very similar for GCO) stacked along [1 0 1] consisting in array of vortices and anti-vortices in the lattice plane containing both $k$-vectors. (f) Details for each Bravais lattice of the spin orientation obtained by combining one or two Fourier components (see text).}
	\label{figDiffraction}
\end{figure}

\subsection{Stabilized magnetic structure}

The starting point of the spin wave modeling described in the previous section is the classical state minimizing the energy of Hamiltonian (\ref{eqham}), considering the interaction set that yields the best agreement between the measured and calculated excitations for both compounds. These ground states were obtained using a real-space iterative minimization method within the rhombohedral magnetic primitive unit cell containing $4\times2^3=32$ spins, which takes into account the possible combination of several propagation vectors (see Appendix E of the Supplemental Material \cite{SupplementalMaterial} for more details). The last stringent test for the relevance of our Hamiltonian was achieved by comparing the single-crystal diffraction Bragg peak intensities obtained from the calculated Fourier components of our GNO and GCO models (see Tables SM3 and SM4 in Supplemental Material \cite{SupplementalMaterial}) to the experimental intensities measured on both compounds using the single-crystal diffractometer D23. We should emphasize that, after adequate scaling of the data (see Appendix A of the Supplemental Material \cite{SupplementalMaterial}) and with the assumption of equipopulated domains, the only adjustable parameter of the fit was the ordered magnetic moment, which was found equal to $M_{\rm Ni}=1.64(4)$ $\mu_\text{B}$ and $M_{\rm Co}=3.23(2)$ $\mu_\text{B}$. The agreement is excellent for GCO (Fig. \ref{figDiffraction}a). For GNO, it is less good, probably due to the presence of two crystallites, but still satisfactory (Figs \ref{figDiffraction}b). The simultaneous description of both the spin waves and the diffraction data from the same Hamiltonian is a strong indication of the robustness of the model. 

Our analysis leads to an original 2-$k$ non-collinear almost-coplanar magnetic structure made of stacking of (anti)-vortices arrays (see Fig. \ref{figDiffraction}e). It is declined in 6 magnetic $k$-domains resulting from all combinations of two $k$-vectors. In each domain, the spin orientation of two of the Bravais lattices are given by these two $k$-vectors whereas the other two are described by only one of the $k$-vectors (see Fig. \ref{figDiffraction}f and Supplemental Material \cite{SupplementalMaterial}). In this structure, all the spins lie close to a plane (perpendicular to [1/3 0 1] in GNO and to [1/6 0 1] in GCO for the domain shown in Fig. \ref{figDiffraction}c,d), which is tilted relative to the plane containing the two $k$-vectors that describe the structure of the domain in question (perpendicular to [1 0 1]).

 In the resulting magnetic structure, the spins whirl around plaquettes forming rows of vortices and of antivortices, both with alternating directions of rotation. This original order results from a compromise between the single-ion Ising-like anisotropy and the competing exchange interactions, especially the strongest FM $J_1$ and AFM $J_3$ ones. This magnetic arrangement can be seen as the flattened version of the 3-$k$ cuboctahedral stack proposed in the theoretical works of refs \cite{Lapa2012,Zhitomirsky2022}, i. e. a 3D periodic arrangement of Bloch-points in its circulating version of Fig. \ref{figStructure}c. The transition from this non-coplanar 3-$k$ structure to this almost coplanar 2-$k$ structure is far from trivial: it does not correspond to the simple extinction of one of the Fourier components, but rather to a transformation of the four Bravais lattices (see Supplemental Material \cite{SupplementalMaterial} for more details). Note that the structure is slightly more flattened in GNO than GCO in line with a less Ising-like single-ion anisotropy in the Ni compound. Nevertheless, both structures belong to the same type IV magnetic space group $C2/c.1_a'[P2/c]$, with a type II point group $2/m.1'$ (see Appendix F of the Supplemental Material \cite{SupplementalMaterial} for more details).

A fit of the data with a single-$k$ structure based on a collinear model is also possible. However, in the calculations, single-$k$ collinear orders are stabilized at a much higher energy than the multi-$k$ structures, as described in the following and in Appendix E of the Supplemental Material \cite{SupplementalMaterial}, which indicates that supplementary ingredients would be required in the Hamiltonian to stabilize such collinear structures.

\begin{figure}[h!]
	\includegraphics[width=1.0\columnwidth]{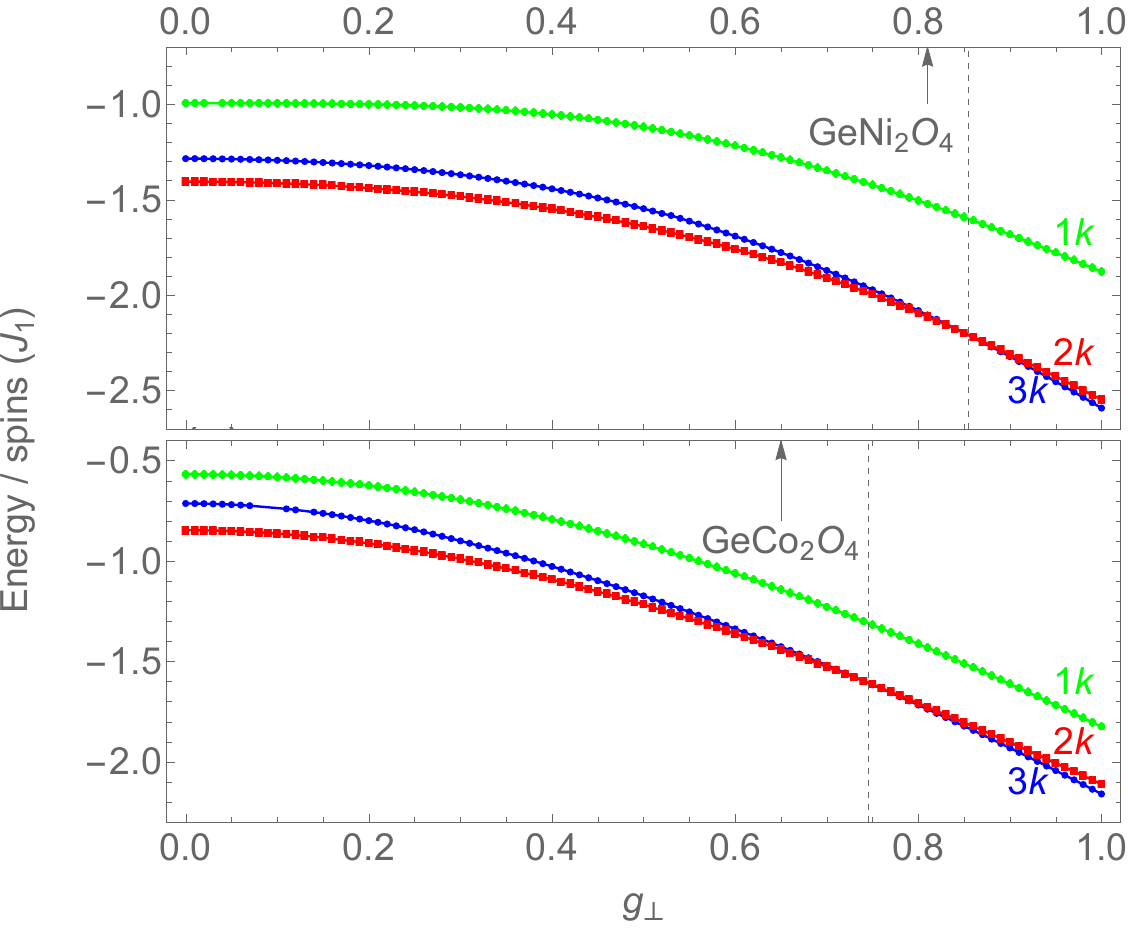}
	\caption{Energies per spins expressed in units of the interaction parameter $J_1$ for the 1-$k$, 2-$k$ and 3-$k$ magnetic structures as a function of the perpendicular component of the anisotropy g$_{\perp}$ (with g$_{\parallel}=1$), and for the best interactions sets given in table \ref{Table1} for GNO (a) and GCO (b). The energies have been calculated using an iterative minimization method \cite{Lapa2012} in the doubled magnetic unit cell containing $4\times 2=8$ (resp. $4\times 2^3=32$) spins for the 1-$k$ (resp. 2-$k$, 3-$k$). The dashed lines indicate the boundary between the regions where the 2-$k$ and the 3-$k$ respectively minimizes the total energy, while the arrows locate the two compounds.}
	\label{figmultikanis}
\end{figure}

\section{Discussion}

Topological interest in non-collinear, non-coplanar spin textures was stimulated by the discovery of magnetic skyrmions \cite{Muhlbauer2009}, then extended to other two- and three-dimensional spin-swirling objects, including various types of vortices and Bloch points/hedgehogs. The status of these complex magnetic objects has evolved from defects to stable entities in periodic arrays. Their role is crucial in the transformation of magnetic matter, the framework for which is provided by topological indicators. The emblematic materials for the stabilization of crystals of such objects are mainly itinerant magnets combining helices with complex competing interactions mediated by conduction electrons. However, the essential ingredients needed to identify potential compounds to host such spin textures are not yet clear.  

In this work, we have shown that short wavelength spin textures can also be observed in insulating antiferromagnets with the simple propagation vector $(\frac 1 2 \frac 1 2 \frac 1 2)$, namely in Ni and Co members of the Ge-spinels. We have considered  exchange interactions up to the sixth neighbors 
and a single-ion anisotropy, which is a very challenging problem to tackle, especially in the context of describing multi-$k$ structures. 
This difficulty explains the contradictory reports regarding the Hamiltonian of these compounds \cite{Diaz2006,Matsuda2008,Pramanik2021,Fabreges2017,Matsuda2011}. 
Indeed, diffraction data or spin wave measurements alone, without a microscopic model that allows us to compare the energies of the 1-$k$ and multi-$k$ states, are insufficient to discriminate between these competing magnetic structures. In the present work, we addressed this problem by first identifying the leading FM $J_1$ and AFM $J_3$ couplings through a study of the magnetic diﬀuse scattering in the correlated paramagnetic state. The Hamiltonian was further refined, including weaker couplings and single-ion anisotropy, through phase diagram calculations, spin wave modeling and confrontation to neutron diﬀraction data.

It is important to notice that the 2-$k$ spin arrangement obtained in our analysis is the one minimizing the Hamiltonian energy (\ref{eqham}) in the cubic system (see previous section and Appendix E of the Supplemental Material \cite{SupplementalMaterial}) and is therefore the best solution for the Ni compound. It bears similarities with the square lattice of whirling objects (e.g. non-coplanar skyrmions and merons or planar vortices) described by a 2-$k$ magnetic structure anticipated theoretically from various Hamiltonians  
\cite{Wang2021,Yi2009,Kato2021,Kato2023,Okumura2020,Park2011,Hayami2018} and observed in the real materials Co$_8$Zn$_9$Mn$_3$ \cite{Yu2018},and GdRu$_2$Al$_2$ \cite{Khanh2020} by Lorentz TEM. Note that, contrary to hexagonal lattices of vortices, the continuity of the spin texture in the square lattice imposes two rotating directions of vortices and anti-vortices. Despite a similar propagation vector in both spinel compounds, in the Co case, a structural distortion has been reported from cubic to lower symmetry \cite{Barton2014} that was not taken into account in our calculations, which were performed in an average high-symmetry cubic structure. This points towards important magnetostructural effects that could drive the material toward a single-$k$ structure as the one described in ref. \cite{Fabreges2017}. In any case, given the small magnitude of the distortion, the GCO Hamiltonian should not deviate too much from that proposed in the present model. Note that this methodology also enabled to identify a multi-$k$ structure describing a complex arrangement of non-collinear spins in the sister compound with Fe \cite{Chaix2026}.

Our work also highlights the role of the single-ion anisotropy, which we find Ising-like in both compounds. The balance between anisotropy and competing interactions produces an interesting change of the magnetic structure. It goes from a 2-$k$ periodic planar arrangement of vortices for Ising-like spins to a 3-$k$ 3D cuboctahedral structure (Bloch-points) for isotropic spins (see Fig. \ref{figmultikanis} and Fig. SM1 in the Appendix B of the Supplemental Material \cite{SupplementalMaterial}). It should be noted that GNO lies very close to the 2$k$-3$k$ phase boundary. One hypothesis, which would deserve further investigations, is that the two temperature transitions represent a shift from the 3-$k$ to the 2-$k$ order, both having energies very close to each other near the transition (while the normalized 1-$k$ state is much higher in energy, because of a mandatory admixture with high energy modes; see Appendix E of the Supplemental Material \cite{SupplementalMaterial} for more details). Such a change in the dimensionality and topology of the spin texture was experimentally observed in Mn(Si$_x$Ge$_{1-x}$) by varying the Si/Ge ratio $x$ \cite{Fujishiro2019}. It was also theoretically predicted in a chiral magnetic metal through the anisotropy of the interactions \cite{Shimizu2021}. In our case, the single-ion anisotropy is the parameter that allows to tune the system between different multi-$k$ structures. At the continuous limit, this would constitute a topological phase transition from a topologically non-trivial hedgehog to a trivial vortex spin structure. Our magnetic order is however short-period and a continuous description must be manipulated very cautiously in the present case. Another way to control the texture and topology of the spin arrangement in these systems could be through the response to epitaxial strain and confinement effects in thin films and heterostructures or by applying a uniaxial pressure. The synthesis of thin films of GeNi$_2$O$_4$ has actually been achieved recently \cite{Vasiukov2021,Indovski}. Another tuning parameter that has been considered is a magnetic field, that produces several spin-reorientations in the Ge-spinels still to be fully understood \cite{Diaz2006,Matsuda2011,Fabreges2017,Basu2020}.

\section{Conclusion}

We have identified in the Ni germanium spinel a frustration-induced 2-$k$ short-period magnetic structure composed of anti-vortices and vortices. This whirling spin texture arises from the single-ion anisotropy-induced flattening of a 3$k$-hedgehog structure predicted for isotropic spins. The same conclusions could apply to the Co sister compound if magnetostructural effects are not a destabilizing factor for magnetic order. 
A similar structure of vortices/antivortices has been observed in metals as well as in frustrated insulators, and also in other fields of physics such as superconductivity \cite{Berdiyorov2006,Campbell2022b,Caggioli2025}, ferroelectricity \cite{Lukyanchuk2024}, fluid dynamics \cite{Chow2003,Gurarie2004}, or optical lattices \cite{Lin2024}, which tends to show that this pattern is rather generic. Our revisitation of these cubic antiferromagnetic compounds, through a coherent methodology combining correlated paramagnetic scattering, phase diagrams, spin waves and diffraction data modeling from the same energy-minimizing Hamiltonian, paves the way for multi-$k$ description of the magnetic structure of numerous frustrated systems, which could broaden the range of materials expected to display spin textures.

\acknowledgments \vspace{-0.1em} We would like to pay tribute to B. F\aa k, who passed away in 2024, and whose contribution to this work was essential. We are grateful to the Institut Laue-Langevin for providing neutron beam time on IN5 (for GNO, \href{https://doi.ill.fr/10.5291/ILL-DATA.TEST-2935}{doi.ill.fr/10.5291/ILL-DATA.TEST-2935} and for GCO, \href{https://doi.ill.fr/10.5291/ILL-DATA.TEST-2662}{doi.ill.fr/10.5291/ILL-DATA.TEST-2662} \& \href{https://doi.ill.fr/10.5291/ILL-DATA.TEST-2965}{doi.ill.fr/10.5291/ILL-DATA.TEST-2965}), on Panther (for GNO and GCO, \href{https://doi.ill.fr/10.5291/ILL-DATA.4-01-1749}{doi.ill.fr/10.5291/ILL-DATA.4-01-1749}) and on D23 (data can be provided on request). We also acknowledge B. Oulladdiaf for allowing us to use Orient Express and P. Fouilloux, J. Halbwachs and O. Meulien for their technical assistance on the ILL instruments. We thank J. Debray for the orientation of some samples. We also acknowledge Laboratoire Léon Brillouin for the beam time on 2T to perform preliminary measurements on GCO. We thank M. Zhitomirsky, Edmond Chan, Fran\c coise Damay and B\'eatrice Grenier for fruitful discussions. Finally, the authors thank the sample providers, Shinichi Ikeda, Shigeo Hara, and Michael K. Crawford.

\bibliographystyle{apsrev4-2}
\bibliography{Spinelle.bib}

\clearpage

\renewcommand{\thefigure}{SM\arabic{figure}}
\renewcommand{\thetable}{SM\arabic{table}}
\renewcommand{\theequation}{SM\arabic{equation}}

\appendix
\section{Refinement of GeNi$_2$O$_4$ (GNO) and GeCo$_2$O$_4$ (GCO) neutron diffraction data in a multi-$k$ description}
\label{ssec:refin}

The magnetic structure factor $\vec{F}_M(\vec{Q})$ is defined by:

\begin{equation}
	\vec{F}_M(\vec{Q})=p\sum_{j=1}^N\vec{m}_j^{\vec{k}} f_j(\vec{Q})e^{-2i\pi\vec{Q}.\vec{r}_j}
\end{equation}	
and
\begin{equation}
	\vec{\mu}_j^l = \sum_{\vec{k}}\vec{m}_j^{\vec{k}} e^{-2i\pi\vec{k}.\vec{R}_l},	
\end{equation}
\\
$p=0.2696\times10^{-12}$ cm/$\mu_\text{B}$ represents the scattering amplitude at $\vec{Q}=0$ for a magnetic moment of 1 $\mu_\text{B}$.
\\
$\vec{\mu}_j^l$ is the total magnetic moment of atom $j$ in cell $l$.
\\
$\vec{m}_j^{\vec{k}}$ is the Fourier component of the magnetic moment $\vec{\mu}_j$ associated to the propagation vector $\vec{k}$.
\\
$f_j(\vec{Q})$ is the magnetic form factor.
\\
$\vec{Q}=h\vec{a}^*+k\vec{b}^*+l\vec{c}^*$, $\vec{r_j}=x_j\vec{a}+y_j\vec{b}+z_j\vec{c}$ and  $\vec{R_l}=u\vec{a}+v\vec{b}+w\vec{c}$.
\\
The analytical approximation of the magnetic form factor is given by : 
\begin{equation}
	f_j(\vec{s}) = j_0(\vec{s}) + \frac{2-g_J}{g_J} j_2( \vec{s} ),
\end{equation}
\\
with $g_J$ the Landé factor, $s = \sin (\theta / \lambda)= Q/(4\pi)$ in \AA$^{-1}$ and
\\
\begin{align*}
	\begin{split} 
		j_0(s)& =A\exp (-a s^2) + B \exp(-b s^2) + C \exp (-c s^2) + D,\\ 
		j_2(s) & = s^2(A2\exp(-a2s^2) + B2\exp(-b2s^2)\\
		& + C2\exp(-c2s^2) + D2.
	\end{split} 
\end{align*}
\\
The coefficients A, a, B, b, C, c, D for l = 0, 2, are tabulated.
\\

\begin{table}[!h]
	\centering
	\begin{tabular}{|p{1.cm}||p{1cm}|p{1cm}|p{1cm}|}
		\hline	 & $g_J$  & $a$ (\AA)  & $\mu$ ($\mu_\text{B}$)  \\
		\hline 	 GNO         & 2 & 8.21 & 1.64(4) \\
		\hline GCO & 3.3 	 & 8.31 & 3.23(2) \\
		\hline
	\end{tabular}
	\caption{Values of the $g_J$ Landé factor, the $a$ lattice parameter and the refined magnetic moment for the 2-$k$ structure in GNO and GCO.  The $g_J$ Landé factor for Co$^{2+}$ was determined from low temperature ESR measurements \cite{Lande} (that accounts for the unquenched angular orbital momentum) and Ni$^{2+}$ was treated in the spin-only approximation.}
	\label{TabMagStrucFactor}
\end{table}

\begin{table}[b!]
	\centering
	\begin{tabular}{@{}c c r r r@{}}
		\toprule
		$j$ & Position & \multicolumn{3}{c}{Local axes} \\
		\midrule
		1 & \multicolumn{1}{c}{(0.25, 0.25, 0)}
		& $1/\sqrt{6}(2$ & -1 &  $1)$ \\
		& 
		& $1/\sqrt{2}(0$ &  1 &  $1)$ \\
		& 
		& $1/\sqrt{3}($-1 & -1  &  $1)$ \\
		\addlinespace
		2 & \multicolumn{1}{c}{(0, 0.25, 0.25)}
		& $1/\sqrt{6}($-2 & -1 & -$1)$ \\
		& 
		& $1/\sqrt{2}(0$ &  1 & -$1)$ \\
		& 
		& $1/\sqrt{3}($1   & -1  & -$1)$ \\
		\addlinespace
		3 & \multicolumn{1}{c}{(0, 0, 0)}
		& $1/\sqrt{6}($-2 &  1 &  $1)$ \\
		& 
		& $1/\sqrt{2}(0$ & -1 &  $1)$ \\
		& 
		& $1/\sqrt{3}($1   &  1  &  $1)$ \\
		\addlinespace
		4 & \multicolumn{1}{c}{(0.25, 0, 0.25)}
		& $1/\sqrt{6}($2  &  1 & -$1)$ \\
		& 
		& $1/\sqrt{2}(0$  & -1 & -$1)$ \\
		& 
		& $1/\sqrt{3}($-1  &  1  & -$1)$ \\
		\bottomrule
	\end{tabular}
	\caption{Position and local axes of the four sublattices $j$.}
	\label{TabSublattices}
\end{table}

\begin{figure}[b!]
	\includegraphics[width=1.0\columnwidth]{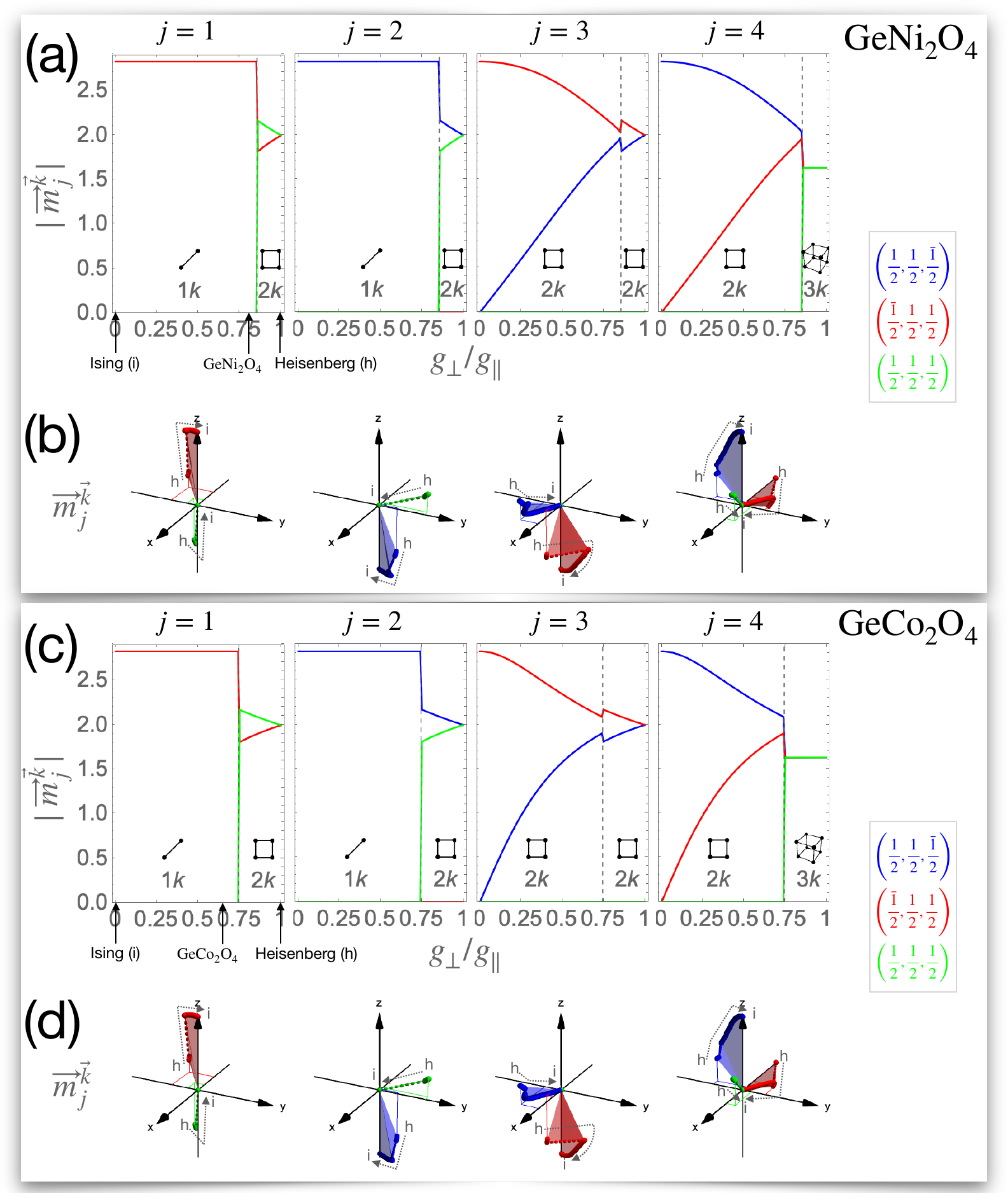}
	\caption{Fourier components of the four sublattices $j=1,2,3,4$ of the magnetic structure of GNO (a,b) and GCO (c,d) as a function of the anisotropy factor g$_{\perp}$/g$_{\parallel}$, calculated using an iterative minimization method \cite{Lapa2012} in the doubled magnetic unit cell containing $4\times 2^3=32$ spins. The norm of the Fourier components are represented in (a,c), while their orientations in each local frames $(\vec{x},\vec{y},\vec{z})_{j}$ (see Table \ref{TabSublattices}) are shown in (b,d). Red, Green and Blue correspond to the propagation vectors $(\frac 1 2 \frac 1 2 \frac {\bar 1} {2})$, $(\frac {\bar 1} {2} \frac 1 2 \frac 1 2)$ and $(\frac  1 2 \frac 1 2 \frac 1 2)$ respectively, and the dashed lines indicate the transition between the 3-$k$ and 2-$k$ phases obtained for the final interaction sets of each compound (see main text).}
	\label{figanisFourier}
\end{figure}

To calculate the magnetic structure factor from our calculated Fourier coefficients, we used the parameters given in the two first columns of Table \ref{TabMagStrucFactor}. We multiplied the calculated Fourier components by the factor $p \mu_\text{B} f(\vec{Q})$. We divided the measured intensities by the scale factor obtained from the nuclear refinement ($I_{obs}=SF_{obs}^2$ with $S$ the scale factor). Finally, we could compare the calculated versus measured squared magnetic structure factor and refine the magnetic moments (last column of Table \ref{TabMagStrucFactor}).

\begin{table*}[!t]
	\begin{equation}
		\begin{array}{c|cccc|cccc}
			\vec{k}  & (\vec{m}_1^{\vec{k}})_{glo} & (\vec{m}_2^{\vec{k}})_{glo} & (\vec{m}_3^{\vec{k}})_{glo} & (\vec{m}_4^{\vec{k}})_{glo} & (\vec{m}_1^{\vec{k}})_{loc} & (\vec{m}_2^{\vec{k}})_{loc} & (\vec{m}_3^{\vec{k}})_{loc} & (\vec{m}_4^{\vec{k}})_{loc} \\
			\hline
			\left(\frac{\bar 1}{2},\frac{\bar 1}{2},\frac{1}{2}\right) & 
			\begin{array}{c}
				0 \\
				0 \\
				0 \\
			\end{array}
			& 
			\begin{array}{c}
				-1.82 \\
				1.91 \\
				1.02 \\
			\end{array}
			& 
			\begin{array}{c}
				-1.19 \\
				1.32 \\
				0.61 \\
			\end{array}
			& 
			\begin{array}{c}
				-1.47 \\
				1.48 \\
				0.32 \\
			\end{array}
			& 
			\begin{array}{c}
				0 \\
				0 \\
				0 \\
			\end{array}
			& 
			\begin{array}{c}
				0.28 \\
				0.63 \\
				-2.74 \\
			\end{array}
			& 
			\begin{array}{c}
				1.76 \\
				-0.5 \\
				0.42 \\
			\end{array}
			& 
			\begin{array}{c}
				-0.73 \\
				-1.28 \\
				1.52 \\
			\end{array}
			\\
			\hline
			\left(\frac{1}{2},\frac{\bar 1}{2},\frac{\bar 1}{2}\right) & 
			\begin{array}{c}
				-1.82 \\
				-1.91 \\
				1.02 \\
			\end{array}
			& 
			\begin{array}{c}
				0 \\
				0 \\
				0 \\
			\end{array}
			& 
			\begin{array}{c}
				-1.47 \\
				-1.48 \\
				0.32 \\
			\end{array}
			& 
			\begin{array}{c}
				-1.19 \\
				-1.32 \\
				0.61 \\
			\end{array}
			& 
			\begin{array}{c}
				-0.28 \\
				-0.63 \\
				2.74 \\
			\end{array}
			& 
			\begin{array}{c}
				0 \\
				0 \\
				0 \\
			\end{array}
			& 
			\begin{array}{c}
				0.73 \\
				1.28 \\
				-1.52 \\
			\end{array}
			& 
			\begin{array}{c}
				-1.76 \\
				0.5 \\
				-0.42 \\
			\end{array}
			\\
		\end{array}
	\end{equation}
	\caption{Example of the Fourier components for GNO for one $2k$-domain. The first column indicates the member of the $k$-star involved. The second and third columns gives the Fourier component coordinates for each of the four Bravais lattice sites (i.e., the four atoms of a tetrahedron, see Table \ref{TabSublattices}) in the global and local frames, respectively. The visualization of the spin direction is given in Figs. 7(c) and e of the main text.}
	\label{Table3}
\end{table*}

\begin{table*}[!tb]
	\begin{equation}
		\begin{array}{c|cccc|cccc}
			\vec{k}  & (\vec{m}_1^{\vec{k}})_{glo} & (\vec{m}_2^{\vec{k}})_{glo} & (\vec{m}_3^{\vec{k}})_{glo} & (\vec{m}_4^{\vec{k}})_{glo} & (\vec{m}_1^{\vec{k}})_{loc} & (\vec{m}_2^{\vec{k}})_{loc} & (\vec{m}_3^{\vec{k}})_{loc} & (\vec{m}_4^{\vec{k}})_{loc} \\
			\hline
			\left(\frac{\bar 1}{2},\frac{\bar 1}{2},\frac{1}{2}\right) & 
			\begin{array}{c}
				0 \\
				0 \\
				0 \\
			\end{array}
			& 
			\begin{array}{c}
				-1.88 \\
				1.93 \\
				0.87 \\
			\end{array}
			& 
			\begin{array}{c}
				-1.21 \\
				1.27 \\
				0.41 \\
			\end{array}
			& 
			\begin{array}{c}
				-1.56 \\
				1.52 \\
				0.06 \\
			\end{array}
			& 
			\begin{array}{c}
				0 \\
				0 \\
				0 \\
			\end{array}
			& 
			\begin{array}{c}
				0.39 \\
				0.75 \\
				-2.7 \\
			\end{array}
			& 
			\begin{array}{c}
				1.68 \\
				-0.6 \\
				0.27 \\
			\end{array}
			& 
			\begin{array}{c}
				-0.68 \\
				-1.12 \\
				1.74 \\
			\end{array}
			\\
			\hline
			\left(\frac{1}{2},\frac{\bar 1}{2},\frac{\bar 1}{2}\right) & 
			\begin{array}{c}
				-1.88 \\
				-1.93 \\
				0.87 \\
			\end{array}
			& 
			\begin{array}{c}
				0 \\
				0 \\
				0 \\
			\end{array}
			& 
			\begin{array}{c}
				-1.56 \\
				-1.52 \\
				0.06 \\
			\end{array}
			& 
			\begin{array}{c}
				-1.21 \\
				-1.27 \\
				0.41 \\
			\end{array}
			& 
			\begin{array}{c}
				-0.39 \\
				-0.75 \\
				2.7 \\
			\end{array}
			& 
			\begin{array}{c}
				0 \\
				0 \\
				0 \\
			\end{array}
			& 
			\begin{array}{c}
				0.68 \\
				1.12 \\
				-1.74 \\
			\end{array}
			& 
			\begin{array}{c}
				-1.68 \\
				0.6 \\
				-0.27 \\
			\end{array}
			\\
		\end{array}
	\end{equation}
	\caption{Example of the Fourier components for GCO for one $2k$-domain. The first column indicates the member of the $k$-star involved. The second and third columns gives the Fourier component coordinates for each of the four Bravais lattice sites (i.e., the four atoms of a tetrahedron, see Table \ref{TabSublattices}) in the global and local frames, respectively. The visualization of the spin arrangement is given in Fig. 7d of the main text.}
	\label{Table2}
\end{table*}

\section{Fourier components of the best models for GNO and GCO 2-$k$ magnetic structure}
\label{ssec:FC}

The Fourier components of the 2-$k$ magnetic structures obtained from the parameters of Table I in the main text for GNO and GCO are given in Tables \ref{Table3} and \ref{Table2}, respectively, in local and global (cubic) frames. The convenience to use local frame in pyrochlore systems is explained in reference \cite{Yan2017} and the positions and local axes of the four Bravais lattice sites are provided in Table \ref{TabSublattices}. Note that, for a given $k$-domain, three atoms of a tetrahedron have non-zero Fourier components for both $k$-vectors, while the latter has only one non-zero Fourier component (see section \ref{ssec:calc} for more details). 

The $g_\perp/g_\parallel$ dependence of the Fourier components is also shown for the GNO and GCO best models in figure \ref{figanisFourier}, which evidences that the 3-$k$ to 2-$k$ transition is not trivial. It does not correspond to a simple extinction of one of the Fourier components, but rather to a different evolution of the four sublattices through the transition. Close to the isotropic case $g_\perp/g_\parallel=1$, the global 3-$k$ structure is composed of one 3-$k$ sublattice and three 2-$k$ sublattices, while the 2-$k$ one is built from two 2-$k$ and two single-$k$ sublattices at lower $g_\perp/g_\parallel$ ratios. In that regime, the norms of the Fourier components of the 2-$k$ sublattices are different: one smoothly tends to zero, such that each sublattice is single-$k$ at $g_\perp/g_\parallel=0$, but the global structure remains described by two propagation vectors (($\frac{\bar1}{2}\frac{1}{2}\frac{1}{2})$ and ($\frac{1}{2}\frac{1}{2}\frac{\bar1}{2})$ in the figure).

\section{Fit of the magnetic excitations}

The magnetic excitations were fitted by one or several gaussian functions:
\begin{equation}
	S(\vec{Q}, \hbar\omega)
	= b_g +
	\sum_{i=0}^{N-1}
	A(\hbar \omega,T)_i \,
	\exp\!\left[
	-\ln(2)\left(\frac{\hbar \omega - \epsilon_i}{\sigma_i}\right)^{2}
	\right],
\end{equation}
where $b_g$ is the background, $
A(\hbar \omega,T) = [1+n(\hbar \omega,T)] \, Z \, \frac{\hbar \omega}{2} \, \frac{1}{\sigma} \sqrt{\frac{\ln 2}{\pi}}
$, $n(\hbar\omega,T) = 1/(e^{\hbar \omega / k_B T}-1)$ is the Bose factor, $Z$ is the weight of the excitations, $\epsilon$ is their energy and $\sigma$ the half width at half maximum.

The energy of the fitted excitations is reported as white dots in Figs. 5 and 6 of the main text, \ref{FigureSM2} and \ref{FigurePramanik}.

\begin{figure}[t!]
	\includegraphics[width=1\columnwidth]{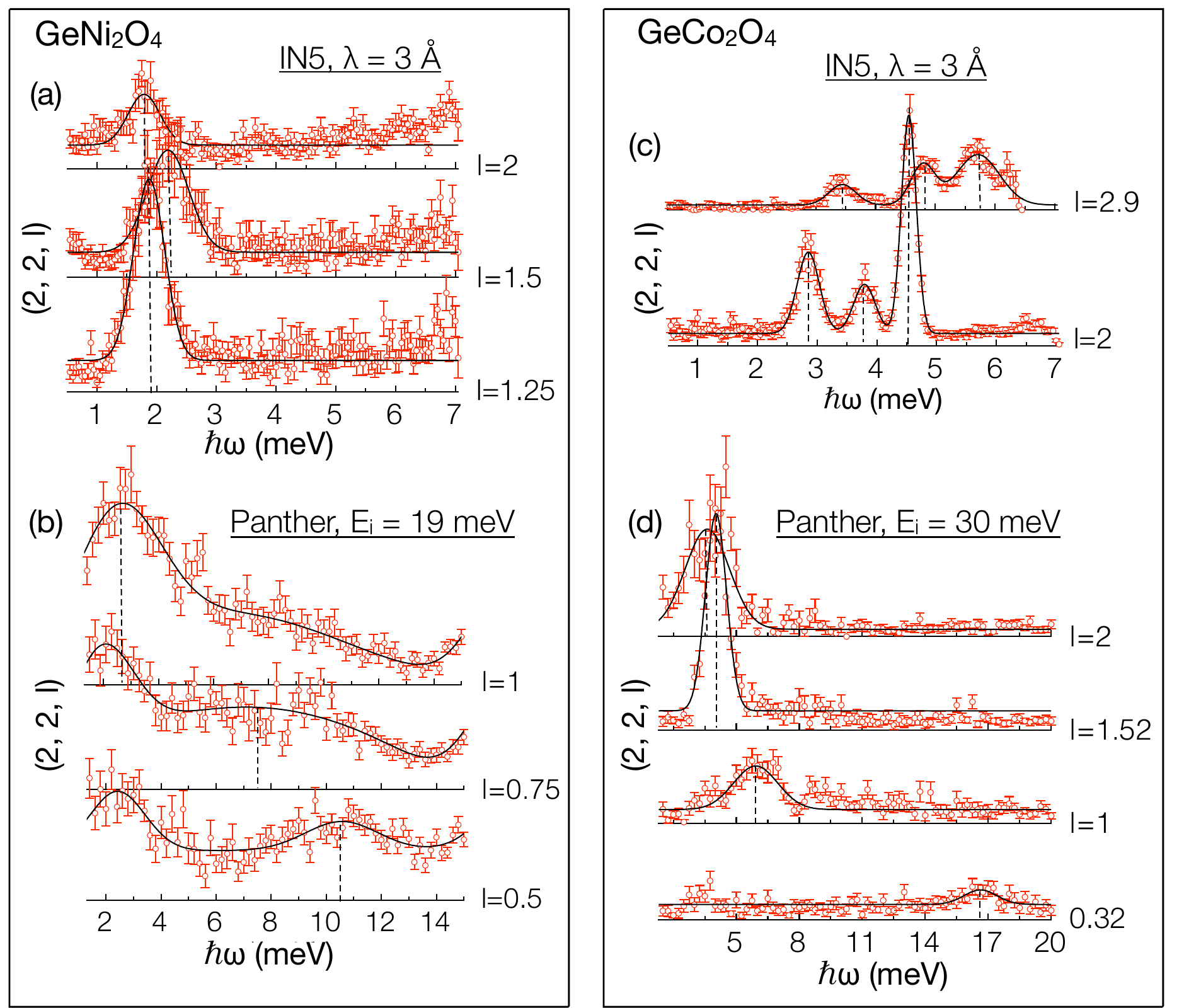}
	\caption{Series of measured constant-$\vec Q$ energy scans along (2, 2, l) (r.l.u) and fits of the excitations (black lines) for GNO (a, b) and GCO (c, d). The vertical dashed lines indicate the fitted energy positions of the experimental spin wave dispersion.}
	\label{FigurePramanik}
\end{figure}

\section{Additional spin waves modeling}
\label{ssec:addSW}

\begin{figure*}[t!]
	\includegraphics[width=2\columnwidth]{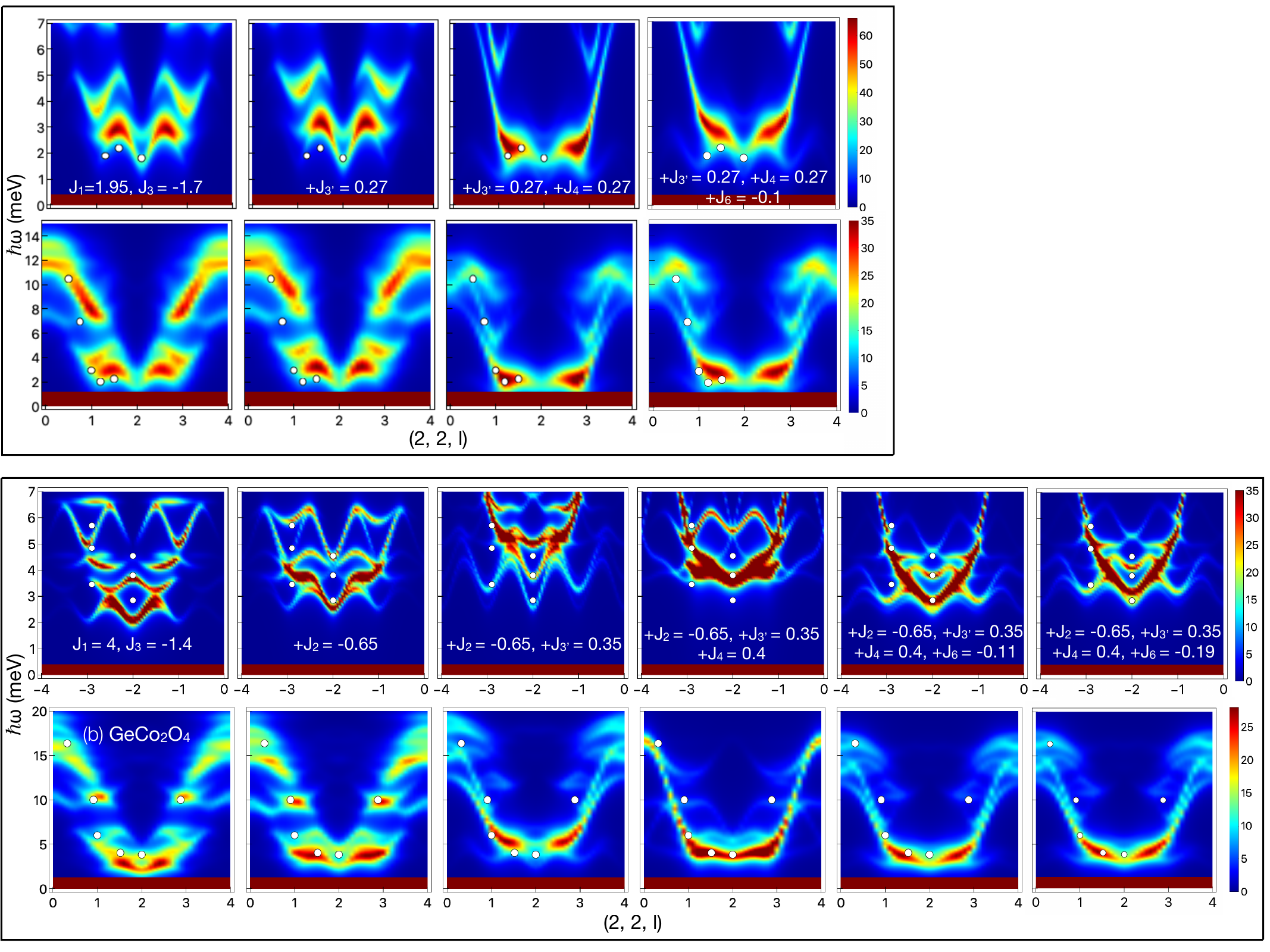}
	\caption{Calculated dynamic structure factor $S(\vec Q,\hbar \omega)$ along (2 2 l) in a) GNO and b) GCO. Influence of the weak interactions $J_{3'}$, $J_4$ and $J_{6}$, as well as $J_{2}$ for GCO, on the shape and spectral weight of the low energy branches.}
	\label{FigureSM2}
\end{figure*}

Further spin waves calculations are presented in this section. In Fig. \ref{FigureSM2}, the high sensitivity of the spin waves to the weak interactions $J'_{3}$, $J_4$ and $J_6$ in GNO, as well as $J_{2}$ for GCO, is illustrated. We note that the model could provide a slightly better description of the measured spin waves with smaller $|J_{6}|$ values (e. g. $J_6$=0 instead of -0.101 for GNO and $J_{6}$ = -0.11 instead of -0.192 for GCO). However, these improved parameter sets do not stabilize the correct $(\frac 1 2 \frac 1 2 \frac 1 2)$ magnetic phase. Figure \ref{FigurePramanik} shows the spin waves calculations based on the model proposed by Pramanik {\it et al.} (parameters in Table I) \cite{Pramanik2021}. The agreement is clearly worse, especially for the low energy excitations.

Note that some small discrepancies between the measured spin waves and the calculated spectra based on our models could be due to additional weak interactions, such as dipolar interactions or anisotropic exchange interactions. Although we did not fully implemented these additional terms, we estimated the strength of the dipolar energy between first neighbors. Expressed per spins, the dipolar energy is equal to 0.018 meV=0.0092$J_1$ and 0.066 meV=0.0165$J_1$ for GNO and GCO respectively. These values are of the order of the energy difference between the 2-$k$ and 3-$k$ states, 0.003$J_1$ and 0.015 $J_1$ respectively for GNO and GCO, and much smaller than the energy difference between the multi-$k$ states to the 1-$k$ one. The dipolar interactions could therefore play a role in the two-step transition observed in GNO.

\section{Details on the calculations }
\label{ssec:calc}

The magnetic phase diagram, as a function of the different exchange parameters, was calculated using the Luttinger-Tisza-Bertaut method \cite{LT,Bertaut,Luttinger_1946}. 
The aim of the method is to determine the ground state of the exchange Hamiltonian (see Eq. (1) in the main text) by formulating the problem in terms of the Fourier Transforms of the spin configuration. For isotropic spins and interactions, diagonalization of the Hamiltonian yields as many eigenmodes as there are Bravais lattices, and the lowest-energy mode (optimal mode) is itself minimized for a set of wavevectors (optimal wavevectors), giving a lower bound for the ground state energy of the system. However, a system containing several Bravais lattices is not guaranteed to reach that energy, since a major difficulty is to achieve normalization of the spins (the so-called ``strong constraint''), which is not imposed during the process. It is indeed sometimes impossible, especially in frustrated systems, to construct a normalized spin state from a single optimal wavevector. In such cases, spin normalization generally requires admixtures of higher energy modes, or preferentially, when possible, a combination of several optimal wavevectors (multi-$k$ spin structure) \cite{Lapa2012}.

In the present case, the eigenmodes of the $(\frac 1 2 \frac 1 2 \frac 1 2)$ phase at the 8 optimal wavevectors of the star $\mathbf{k}_L=\{\frac 1 2\frac 1 2\frac 1 2\}$, are divided into two type of sets, as shown in Fig. \ref{figLTB} : $(i)$ in the 9 low-lying energy modes $i=1,...,9$ (red dot), the Fourier component $(\mathbf{m}^{\mathbf{k}_L}_\alpha)_i$ of sublattice $\alpha$, whose local $\mathbf{z}_\alpha$ component aligns with the optimal wavevector $\mathbf{k}_L$, cancels out, while $(ii)$ in the 3 high energy modes $i=10,...,12$ (black dot), it is the other three spins that cancel out. As a consequence, a $(\frac 1 2 \frac 1 2 \frac 1 2)$ single-$k$ spin normalized state, as the one described in \cite{Fabreges2017}, necessarily emerge from admixture with the high energy modes, and comes at a great energy cost in the present model. On the other hand, generating a normalized spin state from the combination between different optimal wavevectors is achievable in this phase, leading to a ground state energy very close to the lower bound \cite{Lapa2012}. Therefore, during the iterative process that follows, which allows to find the spin configuration minimizing the total energy for a given set of $J_{ij}$, the unit cell is doubled in every direction to allow such multi-$k$ structures, the magnetic unit cell containing $4\times 2^3=32$ spins.

\begin{figure*}[t!]
	\includegraphics[width=2\columnwidth]{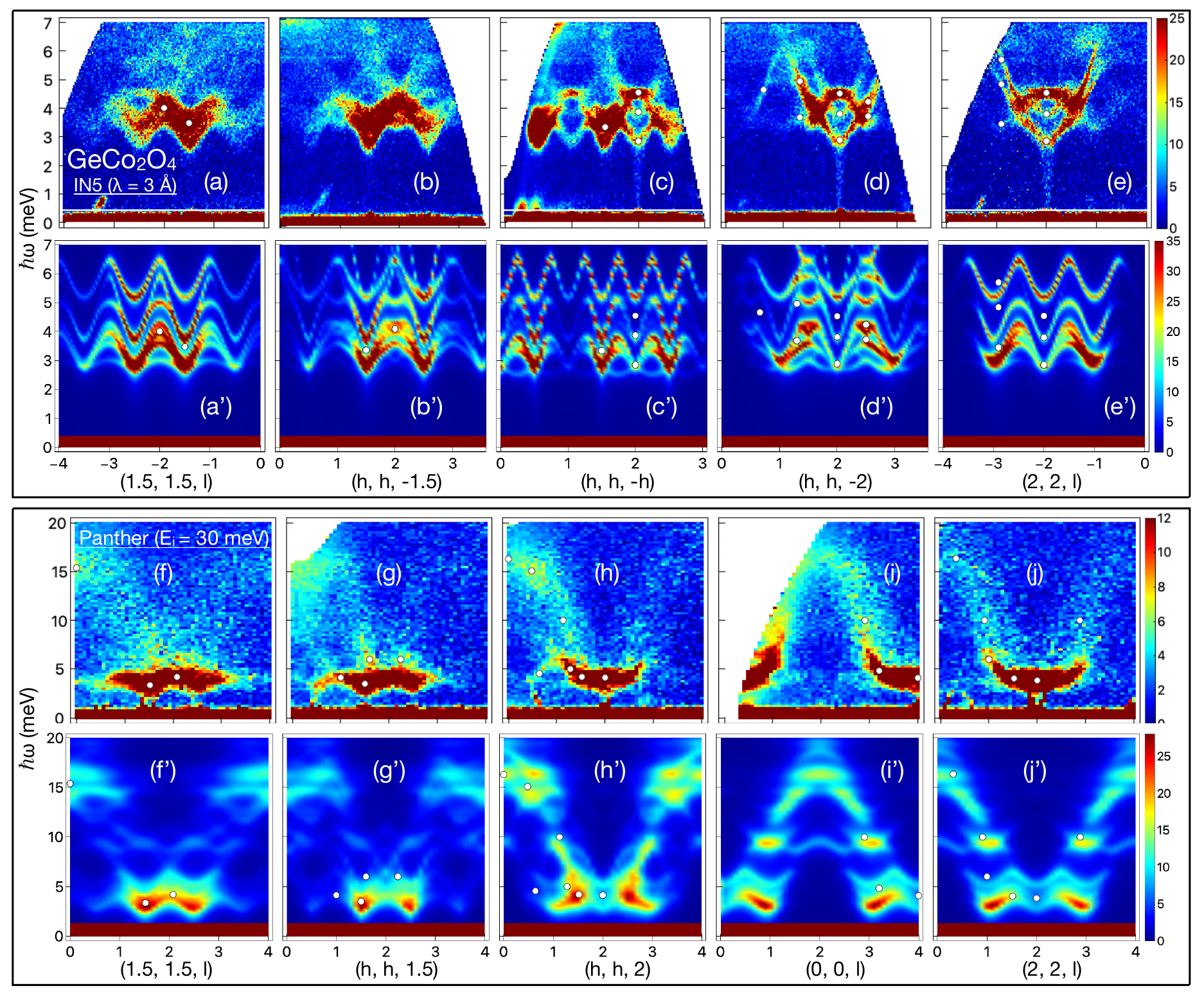}
	\caption{Comparison of the measured dynamic structure factor $S(\vec Q,\hbar \omega)$ in GCO with the one calculated using the Pramanik model (parameters in Table I) \cite{Pramanik2021}.}
	\label{FigurePramanik}
\end{figure*}

\begin{figure}[t!]
	\includegraphics[width=1\columnwidth]{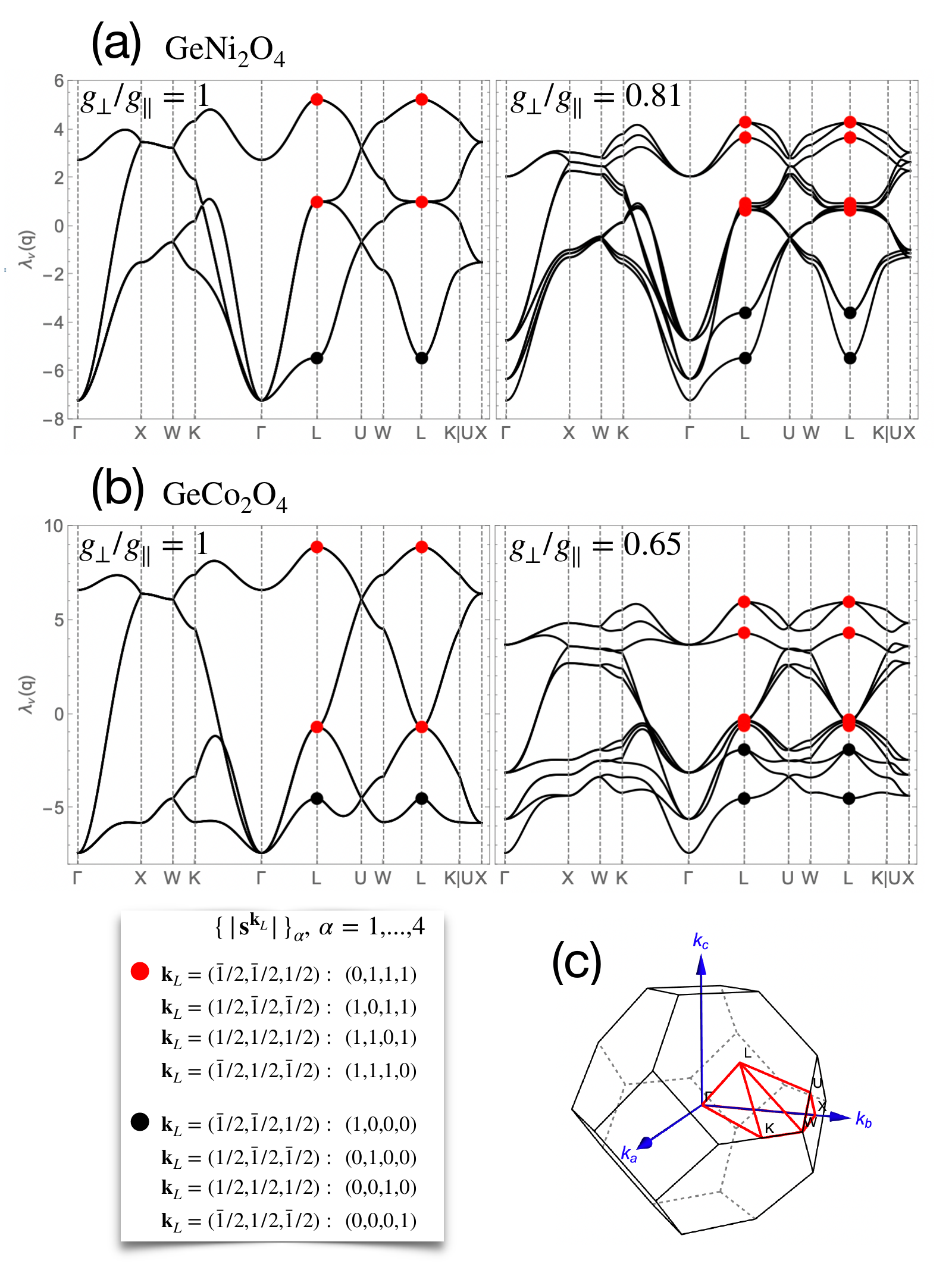}
	\caption{The eigenvalues $\lambda_\nu(q)$ of the matrix $J(q)$ from models of Table I are plotted along a high symmetry route through the Brillouin zone of the face-centered-cubic lattice (c) for GNO (a) and GCO (b). The case of isotropic (resp. anisotropic) spins with 4 (resp. 12) eigenvalues is shown on the left (resp. right). The largest eigenvalue of $J(q)$ is maximized at the $L$ point $q=\{\frac{1}{2}\frac{1}{2}\frac{1}{2}\}$: this corresponds to the lowest energy due to the negative sign in Hamiltonian (Eq. (1) of the main text). Red (resp. black) dots show the eigenmodes that cancel out the spin(s) of one (resp. three) sublattice(s) (see legend and text).
	}
	\label{figLTB}
\end{figure}

The resulting ground state was then used to model the spin wave spectrum via Linear Spin Wave Theory (LSWT) using the Holstein-Primakoff formalism \cite{sylvain}.

Monte Carlo calculations were performed using a hybrid Monte Carlo method with a single-spin-flip Metropolis algorithm. The Monte Carlo method is combined to an integration of the nonlinear equations of motion governing the spin dynamics to obtain the dynamical scattering function $S(\vec{Q}, \hbar\omega)$ \cite{taillefumier}. Simulations were performed above the N\'eel temperature $T_\text{N}$, in the correlated paramagnetic regime, on spin systems of size $16 \times L^3$ with $L=24$. The scattering function was averaged over 200 time-integrated spin configurations to obtain reliable statistics. \\

\section{Magnetic space groups of the multi-$k$ structures}
\label{ssec:MSG}

The Shubnikov groups of the 2-$k$ and 3-$k$ magnetic structures were obtained by identifying the symmetry operations of the paramagnetic space group Fd$\bar3$m.1' that preserve the spin configurations. These results were further verified using the version  7.1.7 of the FINDSYM program within the ISOTROPY Software Suite \cite{findsym,findsymOL}.

\begin{figure}[t!]
	\includegraphics[width=1\columnwidth]{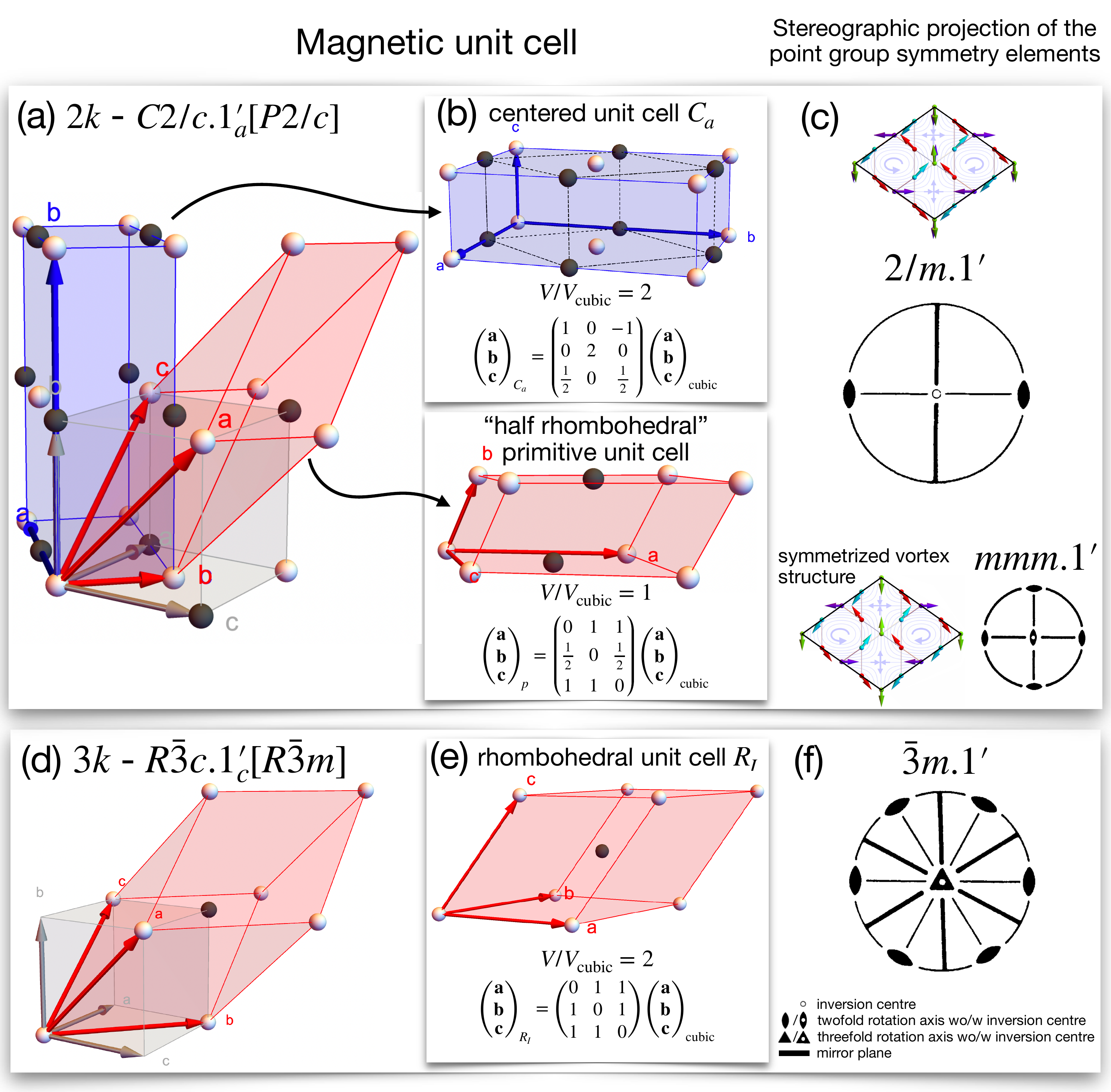}
	\caption{Magnetic space groups of the 2-$k$ (a,b,c) and 3-$k$ (d,e,f) magnetic structures (see main text, Figs. 1 and 7 of the main text): (a,b,d,e) shapes, volumes and relations between the different unit cells.  The black and white dots represent the white and black Bravais lattices connected by antitranslations. Primitive (resp. face-centered) unit cells are shown in red (resp. blue), and the cubic unit cell is represented in gray for comparison. The dotted black lines in the upper panel of (b) represent another possible unit cell. The 2-$k$ primitive unit cell is half of the 3-$k$ one since the structure is doubled in two directions only ($\mathbf{a}$ and $\mathbf{c}$ on the graphic). (c,f) Stereographic projections of the point groups symmetry elements, with graphical symbols taken from \cite{Aroyo2016}. Since the magnetic point groups are gray ones, only the diagrams of the cristallographic point groups are represented.}
	\label{figMSG}
\end{figure}

The 3-$k$ cuboctahedral magnetic structure, with Fourier components given in \cite{Lapa2012}, has a type IV (black-and-white) rhombohedral space group $\mathcal{M}_{3k}=R\bar3c.1_c'[R\bar3m]$ (UNI symbol \cite{Campbell2022,Campbell2024}), with a type II gray point group $\bar3m.1'$ (see Fig. \ref{figMSG} (d,e,f)). Here, $\mathcal{M}_{3k}=\mathcal{D}^S_{3k}+\{1'|\mathbf{t}_{3k}\}\mathcal{D}^S_{3k}$, with $\mathcal{D}^S_{3k}=R\bar3m$ the maximal space subgroup, $\mathbf{t_{3k}}=(\frac{1}{2}\frac{1}{2}\frac{1}{2})$ the antitranslation, and $1'$ the time-reversal group. Its volume is 8 times larger than the paramagnetic (rhombohedral) primitive unit cell, and twice the face-centered cubic one.

In contrast, the 2-$k$ magnetic structure (Fig. \ref{figMSG} (a,b,c)) belongs to a type IV monoclinic $(a,b)$ face-centered space group $\mathcal{M}_{2k}=C2/c.1_a'[P2/c]$, with a type II point group $2/m.1'$. The maximal space group and the antitranslation are respectively given by $\mathcal{D}^S_{2k}=P2/c$, and $\mathbf{t}_{2k}=(\frac{1}{2}00)$, $(0\frac{1}{2}0)$. 
The primitive unit cell can be represented as a rhombohedral unit cell doubled in the two directions of the propagation vectors (see Fig. \ref{figMSG}b, lower panel), and is consequently 4 times larger than the paramagnetic rhombohedral unit cell, and half of the $3k$ rhombohedral magnetic cell (doubled in the three directions) as mentioned above.

It is finally interesting to note that the 2-$k$ magnetic structure determined for the two compounds and shown in the figure 7(c,d,e,f) is very close to a more symmetrical one, with a ``perfect'' vortex-antivortex texture lying in a $\langle1,0,1\rangle$ plane as represented in the inset of figure \ref{figMSG}c. The corresponding space group is $Cmca.1'_a[Pmcm]$ with a gray point group $mmm.1'$, containing two supplemental orthogonal twofold rotation axis (see inset of panel (c)), compared to the original magnetic structure.

\clearpage

\end{document}